\documentclass[10pt,journal,letterpaper,compsoc]{IEEEtran}

\usepackage{cite}
\usepackage{graphicx}
\usepackage{array}
\usepackage{subfigure}
\usepackage{url}
\usepackage{amsmath}
\usepackage{mathrsfs}
\usepackage{amssymb}

\newcommand{\fG}{{\rho_e}}

\newcommand{\fE}{f}

\newcommand{\tsg}{\epsilon_e}
\newcommand{\tsm}{\epsilon_k}

\newcommand{\NOC}{\tau}

\newcommand{\vX}{C}
\newcommand{\vx}{c}

\newcommand{\fpm}{\mathscr{P}}
\newcommand{\vms}{\mathbf{x}}

\newcommand{\ECT}{CT}
\newcommand{\ECP}{CP}
\newcommand{\EPC}{PR}

\newcommand{\vp}{p}

\newcommand{\vm}{M}

\newcommand{\sd}{\mathbf{n}}    

\newcommand{\Eq}[1]{(\ref{#1})}
\newcommand{\Fig}[1]{Fig. \ref{#1}}

\newcommand{\Sec}[1]{Section \ref{#1}}
\newcommand{\vnu}{\mathcal{O}}
\newcommand{\vxx}{\mathbf{x}}

\newcommand{\vxv}{\mathbf{v}}

\begin{document}
%
\title{Reconstructing 3D Motion Trajectory of Large Swarm of Flying Objects}

\author{Danping~Zou~ and~Yan~Qiu~Chen
\IEEEcompsocitemizethanks{\IEEEcompsocthanksitem School of Computer Science, Fudan University, Shanghai, China.\protect\\
E-mail: [dpzou/chenyq]@fudan.edu.cn}
\thanks{Preliminary results of the research work described in this paper ware presented at ICCV 2009. This draft was finished at Oct,2011.}}

\IEEEcompsoctitleabstractindextext{%
\begin{abstract}
  This paper addresses the problem of reconstructing the motion trajectories of the individuals in a large collection of flying objects using two temporally
synchronized and geometrically calibrated cameras. The 3D trajectory
reconstruction problem involves two challenging tasks - stereo
matching and temporal tracking. Existing methods separate the two
and process them one at a time sequentially, and suffer from
frequent irresolvable ambiguities in stereo matching and in
tracking. We unify the two tasks, and propose an optimization approach to
solving stereo matching and temporal tracking simultaneously. It
treats 3D trajectory acquisition problem as selecting appropriate
stereo correspondence out of all possible ones for each object via
minimizing a cost function. Experiment results show that the
proposed method offers significant performance advantage over existing
approaches. The proposed method has successfully been applied to reconstruct 3D motion
trajectories of hundreds of simultaneously flying fruit flies (Drosophila Melanogaster), which could
facilitate the study the insect's  collective behavior.
\end{abstract}

}

\maketitle

\IEEEdisplaynotcompsoctitleabstractindextext
\IEEEpeerreviewmaketitle
\section{Introduction}
\IEEEPARstart{A}{prevalent} phenomenon in nature is aggregations of objects moving in a 3D space,
such as insect warms, bird flocks, and fish schools.
These subjects tend to create complex dynamic behavior. Birds and
fish gather in vast numbers, keeping some sort of cohesion in movement and
creating fascinating patterns\cite{emlen1952flocking}\cite{pitcher1993behaviour}.
Migrating butterflies fly within a bounded layer above the ground
towards one direction\cite{ taylor1974insect}. Bats simultaneously
emerge in great number from cave and soar in the sky at dusk
\cite{reichard2009evening}. Studying the collective behavior of
these animal aggregations is of great value to a wide range of
fields, including evolutionary biology, 
artificial intelligence\cite{ kennedy2006swarm }, computer graphics
\cite{reynolds1987flocks }, control
theory\cite{jadbabaie2003coordination},engineering\cite{de2006swarmav},
economics\cite{cont2000herd} and social sciences \cite{
helbing2000simulating },

An effective way to study the behavior of animal aggregations is 
through accurately measuring the 3D motion pattern of these subjects, or in other words, 
measuring how the 3D location of each individual varies with time.
 Quantitative analysis can then be done to aid in discovering and explaning the underlying behavior patterns of the subjects. Since there had to date not been effective methods to measure the trajectory of each individual in a large swarm, visual inspection,
 instead of quantitative analysis, is used to make conjectures.  Although some mathematical models such as \cite{reynolds1987flocks}\cite{vicsek1995novel}\cite{couzin2002collective},  have successfully simulated the life-like collective behavior by assuming some simple rules, without accurate measurement data of real-world cases, as stated in \cite{parrish1999complexity}, it cannot be verified that the living systems actually follow these rules.  So the lack of effective methods to measure the 3D motion trajectories of flying object aggregations has become a bottleneck of the current research on animal collective behavior.

A feasible way to measure the 3D motion
trajectories of object aggregations is through using multiple cameras. It has several advantages
over the sensor-based method in which some positioning and wireless communication devices are mounted
on the subjects\cite{nagy2010hierarchical}. The vision-based method
is able to accurately measure a large number of objects at a low cost and does
not affect the behavior of the subjects.

The vision-based method to recover the
time-varying 3D coordinates of the objects involves two tasks, namely, stereo matching -
establishing stereo correspondences across views, and tracking -
finding motion correspondences for each object. In the case of a large number of visually identical objects flying in a 3D space, both tasks are challenging since no appearance cue (such as color, texture, and shape) is available to distinguish the individuals. There has been research on the problem but no
satisfactory solution has been published.

\begin{figure}[ht]
\centering
\includegraphics[width =
0.22\textwidth]{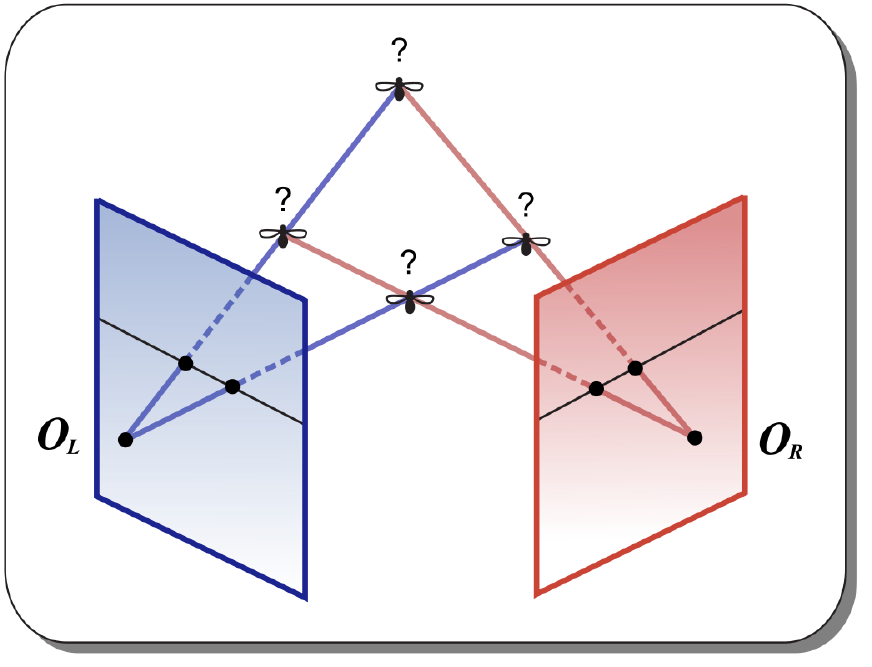}
\caption{ The object on the left image has
multiple matching candidates on the right image satisfying the epipolar
constraint resulting in stereo matching ambiguity.} \label{fig:epipolar_ambiguity}
\end{figure}
Existing related approaches are generally
found in the applications of multi-camera tracking, such as tracking
pedestrians or feature points in multiple cameras, which can be
classified into two major groups. The first category of methods establish stereo correspondences first frame by frame to reconstruct 3D locations of the objects.  The
3D locations corresponding to the same object are then
temporally associated to yield 3D trajectories
\cite{kasagi1991ptt,malik1993ptv,pereira2006tfp,grover2008fly}. This
kind of methods work well when the number of objects is small or
each object carries sufficient distinctive visual appearance that can be used to
identify itself from the others. To deal with a large group of objects
containing many visually identical objects, problems arise - multiple
matching candidates are often present in stereo matching (see
\Fig{fig:epipolar_ambiguity}). It is impossible to identify which
candidate is the genuine stereo correspondence,  as the objects are
visually indistinguishable. The stereo matching ambiguity happens so
frequently that it makes the 3D coordinates reconstructed in the
first step  unreliable.
 Once the stereo matching
ambiguity is incorrectly resolved, the final result will be greatly
deteriorated by the incorrect 3D locations.

\begin{figure}[ht]
\centering
\includegraphics[width =0.22\textwidth]{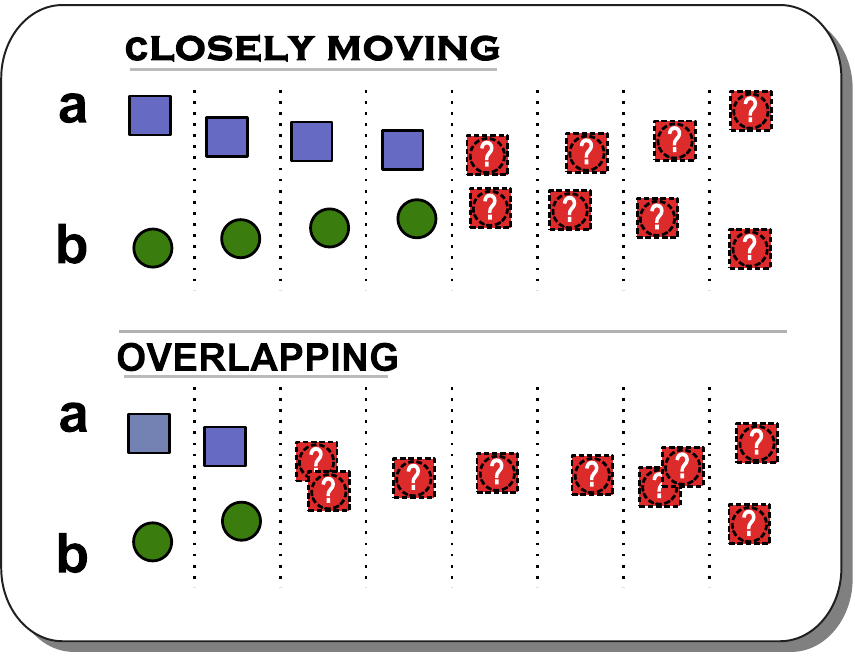}
\caption{Objects $a$ and $b$ overlap on the 2D image plane, leading to tracking
ambiguity.} \label{fig:tracking_ambiguity}
\end{figure}
Approaches of the second category
\cite{du2007rem,guezennec1994afa,engelmann1998phf} track objects on each 2D image plane first. The
resulting 2D tracks are then matched across views using  
inter-camera geometry and motion clues\cite{du2007rem}.  After
that, 3D motion trajectories are reconstructed from the matched
2D tracks. The motivation behind this kind of methods is to utilize
motions of objects on image planes to resolve single-frame
matching ambiguity. This strategy is effective if the objects
seldom overlap one another on image planes when the 2D motion
trajectories of objects can be reliably acquired through tracking.
For a dynamic particle swarm containing many objects flying in
the 3D pace, the images of these objects inevitably overlap at times. This poses a difficult situation for
tracking (see \Fig{fig:tracking_ambiguity}),  which likely causes
failure in obtaining correct 2D tracks. Once 2D tracks are
incorrect,  matching the false tracks will result in severely corrupted 3D trajectories.

For the existing two types of methods, stereo matching and tracking are  separated as two successive stages, where the first stage
has a decisive influence on the final result. The difficulty is no
matter which one is chosen as the first stage, 
such methods tend to produce poor performance for particle swarms. One naturally argues that stereo matching
and tracking are in fact interwind in the problem of 3D trajectory
reconstruction: stereo matching can be done much easier  if tracking
results are known, while tracking on image planes can also be
facilitated if stereo correspondences have been established at the
current time step. A question arises : Is there a unified approach in
which stereo matching and tracking are simultaneously
performed to optimize the performance?

\begin{figure*}
\centering
\includegraphics[width = 0.75\textwidth]{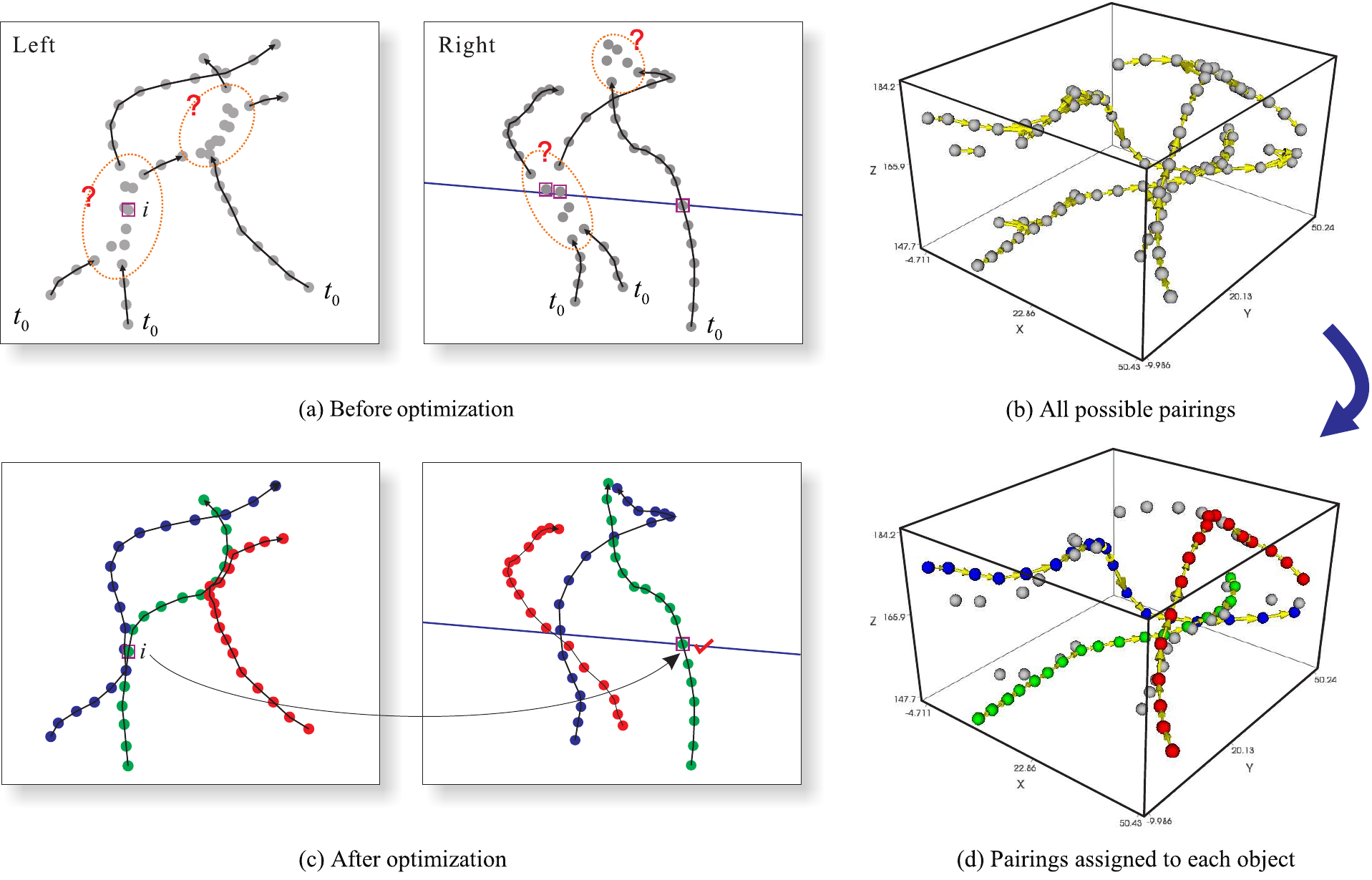}
\caption{An example illustrating the
capability of the proposed method. In (a), tracking and stereo
matching ambiguities both exist,  e.g. the object $i$ has multiple
stereo matches in the right view and it also temporally corresponds
to two possible positions at the next frame. After optimization,
pairing sequences are obtained for each of the three objects (marked
by red,green and blue) as shown in (d) and both ambiguities are
resolved as shown in (c). } \label{fig:simple_example}
\end{figure*}

The answer is affirmative. We propose an optimization approach to
reconstructing the 3D motion trajectories of flying object
aggregations from two temporally synchronized and geometrically
calibrated cameras, which unifies stereo matching and
tracking into a whole.  A key idea of
this method is to treat each stereo pair of image objects as a
\emph{pairing} and convert the trajectory reconstruction problem into a
process of selecting pairing sequence for each object. Through
minimizing a cost function that incorporates cues from epipolar geometry,
motion, and one-to-one match preference, the optimum pairing sequences
of corresponding objects are obtained. The 3D motion trajectories
are finally reconstructed by triangulation from the paired image
locations. Both simulated and real-world experiments show 
remarkable performance of the proposed method.  An illustrative
example of the proposed method is given in \Fig{fig:simple_example}.

The proposed method is able to deal with a large swarm of
objects flying in the scene, and accommodate various kinetic models
according to different situations. By adopting kinetic models on 2D
image planes, the proposed method can also work in the case where
the cameras are weakly calibrated - only the fundamental matrix
between two cameras is known. The contribution of this paper is two
fold:
\begin{itemize}
\item The paper presents a cost
minimization method to solve the stereo matching and tracking
in a unified manner, which according to our survey is the first effective method for
reconstructing the 3D motion trajectories of flying particle-like
objects from two calibrated and synchronized cameras
\item To our best knowledge, this is the first time  that the 3D motion
trajectories of a swarm of hundreds of flying fruit flies are obtained, which
could facilitate the study of 
the insect's collective flight behavior.
\end{itemize}

The remainder of this paper will discuss related work in Section II.
The 3D trajectory acquisition problem is formulated in Section III and
a cost minimization approach is proposed in Section IV. Next, a
sampling method based on solving assignment ranking problem is
presented to optimize the objective function in Section V.
Section VI discusses experimental results on both simulated and real-world swarms.
The conclusion is drawn in Section VII.

\section{Related work}
\subsection{Multi-object tracking techniques}
Numerous approaches have been developed for tracking multiple
objects, including Multpile Hypothesis Tracking(MHT)\cite{reid1979atm}, Joint Probabilistic Data Association Filter(JPDAF)
\cite{fortmann1983stm}, Greedy Assignment\cite{veenman2001rmc}, and
Particle filter\cite{gordon1993nan}. Two issues need to be
addressed in multiple object tracking. The first one, known as data
association, is to establish a mapping between measurements (e.g.,
image blobs) and objects at each time step. The second one is to
estimate the motion states from the identified measurements for each
object. For MHT\cite{reid1979atm}, all feasible mappings between
objects and measurements are enumerated and are thought of as a set
of hypotheses.
 Probability of each hypothesis is propagated from prior hypothesis at previous time step. The associations are finally
 determined
 by selecting the best hypothesis with maximum probability over time.
JPDAF \cite{fortmann1983stm} uses a similar way to evaluate the
probability of each hypothesis, but instead of finding best
hypothesis, it computes expectation of the motion state of objects
over all hypotheses.

Both MHT and JPDAF use brute-force enumeration to generate legal
hypotheses, leading to their incapability of coping with a large
number of objects due to excessive computational complexity.
To efficiently generate hypotheses, the authors of
\cite{danchick1993fmf}\cite{cox1996eir} regard an association as a
bipartite graph match,
 and then find the k-best hypotheses by recursively solving the assignment problem. Sometimes only the best hypothesis is obtained through one-step bipartite graph match, as the greedy optimal assignment (GOA) tracker \cite{veenman2001rmc} does.
Although those bipartite-graph matching based algorithms 
significantly improve computational efficiency,
 their generated hypotheses are restrictive because they assume that a measurement can only be associated with no more
 than one object and
vice verse. This assumption does not always hold in practice, e.g., a
measurement on image plane could correspond to several objects
as a result of occlusion.

Another trend of methods is based on particle filter. The particle
filter\cite{gordon1993nan}, approximating the probability density of
motion state of an object conditioned on measurements by a set of
weighted particles, is able to cope with tracking problem with
non-linear measurement models and non-Gaussian noise where no
analytic expression (closed formula) for probability density of motion
state can be assumed. To solve the multi-object tracking problem,
the particle-filter based methods collect all individual motion
states into a single state variable
 and directly sample the probability densities of the joint motion state over time \cite{hue2002tmo}.
 Unfortunately, the straightforward implementation of the joint particle filter by Sample Importance Resampling
suffers from exponential complexity in the number of objects.
 Therefore, Markov Chain Monte Carlo (MCMC) technique is adopted to sample the high dimensional probability density of
 joint state \cite{maccormick2000pep}.
 Whereas, directly sampling in the continuous state space still requires extremely high computational cost and could
 fail in coping with a large number of objects.
These existing multi-object tracking methods only focus on
tracking objects on monocular video sequences. They are not capable
of tracking the 3D positions of moving objects from stereo video
sequences, since there is an additional issue in addtion to tracking -
stereo matching.

\subsection{Reconstructing 3D motion trajectories of multiple objects}
Most existing methods for recovering 3D motion trajectories
of multiple objects apply techniques of multi-object tracking and
stereo matching sequentially. The strategy of matching 2D
tracks obtained through tracking on 2D image plane has been applied
to reconstruct 3D trajectories of identical objects in binocular
stereo \cite{du2007rem,guezennec1994afa,engelmann1998phf}. The 2D
tracks are however, not guaranteed  to be correct,
because objects on image planes may frequently overlap.   Du et al.
\cite{du2007rem} suggested stop tracking at  the time when
interaction happens, and establish correspondence only for these
partial 2D track segments. However, the common time span between
segments could be too short to resolve stereo matching ambiguity. Although some stitching strategies can be applied, the
resultant 3D trajectories can still be severely broken due to
lengthy interactions as shown in \Fig{fig:2d_flies}.

A few methods incorporate tracking and stereo matching to generate
more reliable results.  Willneff et al. \cite{willneff2002nst}
proposed a spatial-temporal matching algorithm using motion
prediction to reduce stereo matching ambiguities. Unfortunately, the
algorithm would fail when stereo matching ambiguities become severe,
since generating reliable prediction highly depends on correct 3D
locations in the previous frame.

\subsection{Animal and insect tracking for behavior research}
Tracking animals  using video sequences to
facilitate the study of animal behavior, has increasingly attracted
researchers in the computer vision and biology communities.  Z.
Khan et al. \cite{khan2005mbp} tracked 2D trajectories of dozens of
ants through MCMC-based particle filter. A. veeraraghavan et al.
\cite{veeraraghavan2008shape} track a bee dancing in a beehive by
combining motion model and shape model to simultaneously obtain its
2D motion trajectory and shape changes.
 Authors in \cite{tweed2002tracking} obtained both trajectories and skeletons of  flying birds through tracking on the
 image plane. In
 \cite{hirsh}, a real-time system of tracking flying bats in a single view was
 developed. A single-camera system was presented in \cite{branson2009high} to track and classify the behaviors of about
 $50$ fruit flies
 moving on the 2D plane.

 Biologists are not satisfied with acquiring only 2D
motion trajectories for the study of subjects that fly or swim in
3D space. As early as in the 80's of the last century, researchers had
attempted to obtain 3D trajectories of flying fruit flies
\cite{buelthoff1980daf}, house flies \cite{wehrhahn1982tac}, and
bats \cite{rayner1985tdr} using
two film cameras. In recent years, multiple digital video cameras
are used to reconstruct 3D motion trajectories of fruit flies 
\cite{grover2008fly} \cite{dickinson2008}. But they can only
deal with a few subjects by employing the existing techniques
in computer vision. In \cite{ballerini2008interaction}, authors
obtained the 3D positions of individuals in a starling flock by
stereo matching, they however did not maintain their identities over
time. The authors  of \cite{wu2009tracking} measured 3D motion
trajectories of flying bats from three calibrated cameras by solving
multiple-dimensional assignment problem based on the epipolar
constraint at a single video frame. But they presented 3D
trajectories of only about ten flying bats in the video result. Recently, Straw
et al. \cite{straw2010multi} proposed a real-time tracking system with eleven cameras to analyze the effects of 
visual contrast on the flight performance of fruit flies, but their system can only track a few insects.

\section{Problem statement}
This paper studies the problem of reconstructing the 3D motion trajectory of each individual
in a swarm of flying particle-like objects from multiple video
sequences captured from different viewing directions. We focus on using two cameras
 due to its practical advantage of lower costs and easier deployment.  It is not difficult to extend 
the proposed idea to the multiple camera case.

Suppose there are $N$ objects flying in a 3D space. Two temporally synchronized and geometrically calibrated cameras capture 
the scene at time instances, $t=1,2,\ldots,T$, producing two video sequences. 
The target blobs in  the image can be detected and located by object detection techniques  \cite{elgammal2000non} \cite{okuma2004boosted}. We denote the detected blobs at time step $t$ by a set $\vm_t$.  The problem of trajectory reconstruction is to compute the time-varying 3D locations for each object from the observations - blobs detected from stereo video sequences, $\vm_1,\vm_2,\ldots,\vm_T$.

\section{Method}
The trajectory reconstruction problem faces two challenges. One lies in the temporal domain, which is
to establish motion correspondence between video frames.  The other lies in the spatial domain, which is to find stereo
correspondence across views.  As discussed earlier (see \Fig{fig:epipolar_ambiguity} and \Fig{fig:tracking_ambiguity}) ,
the two issues, if addressed separately, will cause ambiguities that are difficult to resolve.  We
propose a novel method in this paper that combines the two issues into a single problem and solve it by
optimizing a unified objective function.

A key concept of the proposed method is \lq pairing\rq.  A pairing denotes a potential stereo correspondence,
where the two paired image blobs may correspond to the same object. Denoting the blobs detected at respective views by $\vm^{[1]}_t,\vm^{[2]}_t$, the possible pairings 
at the time step $t$ is given by $\vm^{[1]}_t\times \vm^{[2]}_t$.  Each pairing is related to a 3D location by triangulating
the centers of the two blobs. The trajectory reconstruction problem therefore becomes selecting the genuine pairings for
each object over time.

In other words, the task is to select a sequence of pairings for each object.  We name the
pairings selected at each time step as a \emph{configuration}, denoted by  $\vX_t =
(\vx^1_t,\ldots,\vx^n_t,\ldots,\vx^N_t)$. The problem we need to solve is to find out a sequence of configurations,
$\vX_{1:T} = (\vX_1,\ldots,\vX_T)$ that best explain the target blobs recorded during the capturing process.
To produce reasonable result, some knowledge is required to evaluate the quality of a given sequence of configurations.
Define the evaluation function as $f(\cdot)$, which should incorporate all aspects of knowledge about what a good
configuration sequence should possess.   Three cues are used. The first is epipolar constraint, which
describes how a pairing geometrically fits as a correct stereo correspondence.  The second is one-to-one match
tendency. It is based on the observation that the majority of blobs have only one corresponding blob in the other view.
The last is the knowledge about the motion of the objects.  It describes how objects move and is used to ensure the
reconstructed trajectories are consistent with the kinetic model of the objects. 
We will provide a detailed description about how these cues are mathematically formulated and incorporated into the
evaluation function $f(\cdot)$ in the following sections.

After that,  the trajectory reconstruction problem becomes an optimization problem, that is, to find an optimum
configuration sequence
$\vX^*_{1:T}$ that minimizes $\fE(\vX_{1:T})$:
\begin{equation}
\vX^*_{1:T}=\arg\min_{\vX_{1:T}}\fE(\vX_{1:T}).
\end{equation}
When an optimum configuration sequence is obtained, the 3D motion trajectories can be computed by stereo
triangulation for each pairing. 

The number of objects present in the scene is usually a variable of the time, which
makes the problem more challenging.  In \cite{ khan2005mbp }, a variable is introduced to indicate the set of
visible objects at each time step, leading to a very complicated algorithm where the dimension of state space varies
with iterations.  Here, we use a dummy pairing $\vnu$ to represent the absent state of objects.  When
objects have not shown up in the scene or have left the scene, they are simply assigned with $\vnu$. The advantage of
this solution is that we can keep the dimension of the state space constant and do not have to resort to a complex algorithm
that jumps among spaces of different dimensions. The drawback is the pairing sequences extracted from the same
dimension of the configuration variable separated by the $\vnu$ may not correspond to the same physical object.
We can however treat these pairing sequences as belonging to different objects, since it is difficult to know
whether an object that newly appears has shown up in the scene before.

The dimension of the configuration, $N$, can be determined by examining the maximum number of objects present in the
scene throughout the time.  The number of objects is usually estimated from the detected blobs in the video sequences.  To account
for the false detections and noises, $N$ is can be set slightly larger than the estimated number of objects.

\subsection{Epipolar constraint}
If the blobs in different views correspond to the same 3D object,
they will satisfy the epipolar constraint. This is an important cue
for judging how likely a pair of blobs is a genuine stereo
correspondence. 

\begin{figure}[ht] \centering
\includegraphics[width=0.2\textwidth]{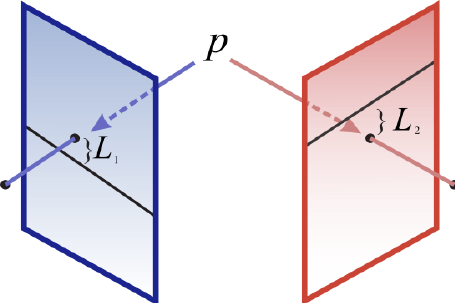}
\caption{The epipolar cost of a pairing $p$ is defined as $\fG(p)
=(L_1+L_2)/2$.} \label{fig:geometric_constraint}
\end{figure}

This implies that the blobs closer to epipolar lines of each other
have higher likelihood of corresponding to the same object. Denoting
a non-dummy pairing by $\vp$, we define the cost of the pairing,
$\vp$, being assigned to the object $n$ as
\begin{equation}\label{eq:epi_cost_single}
 \fE_e(\vx^n_t) = \fG(\vp),
\end{equation}
where $\fG(\vp)$ represents  the average distance between the blob
centroids and their respective epipolar lines as shown in
\Fig{fig:geometric_constraint}.

Suppose $N$ objects appear in the scene during
video capture.  Let $N^*_t \subset \{1,\ldots,N\}$ be the set of
objects assigned with non-dummy pairings in the current
configuration $\vX_t$. The total cost of the configuration
based on epipolar constraint is given by
\begin{equation}
\fE_E(\vX_t) = \frac{1}{|N^*_t|}\sum_{n \in N^*_t} \fE_e(\vx^n_t).
\label{eq:epipolar_cost}
\end{equation}
To rule out apparent false pairings ,
 we set $\fE_e(\vx^n_t) = \infty$, if
$\fG(\vx^n_t)$ is greater than a preset threshold $\tsg$.

\subsection{One-to-one match tendency}\label{sec:one_to_one_match}
Apart from the epipolar constraint, we also measure the level of
consistency between the current configuration $\vX_t$ and observed blobs
by using a criterion based on the following observations: 1) Cameras can be well placed so that they capture most objects
simultaneously; 2) at each time step, the proportion of overlapping
image blobs on the 2D image plane is relatively small if the
density of flying objects is moderate. It indicates that a blob
tends to be related to only one object in the scene. This cue also implies
that each blob most likely corresponds to only one blob in
the other view.

\begin{figure}[ht]
\centering
\includegraphics[width=0.35\textwidth]{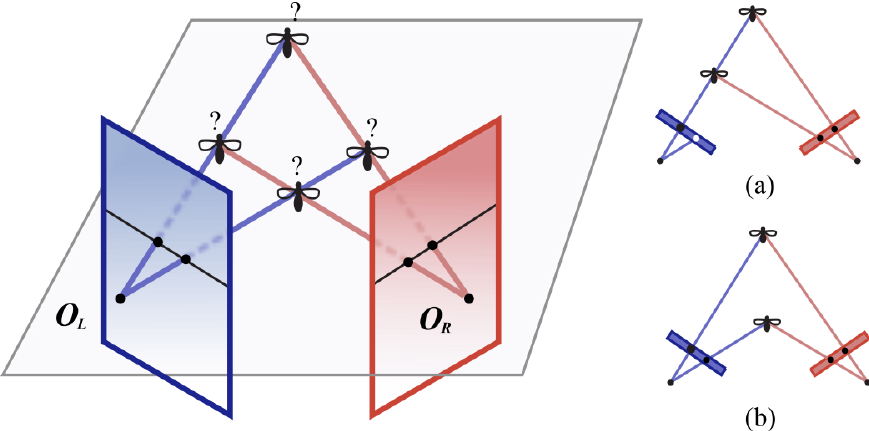}
\caption{ The configuration in (b) is evaluated as more desirable
than the one in (a). Because in (a),repeat assignments and missing
assignments make the matching costs high, while in (b), blobs have
established one-to-one mapping between views, leading to low
matching costs.} \label{fig:consistency}
\end{figure}

Noticing the tendency of one-to-one match between blobs in different
views, we propose two cost functions to penalize missing assignments
and duplicate assignments for detected blobs. Given a configuration
$\vX_t$, for a blob $i \in \vm_t$, the number of its corresponding
blobs in the other view is denoted by $\NOC(i,\vX_t)$. The two cost
functions are defined as
\begin{equation}
\begin{split}
\fE_{C_1}(\vX_t) =& \frac{1}{|\vm_t|} \sum_{i \in \vm^{-}_t}[1-\NOC(i,\vX_t)]\\
\fE_{C_2}(\vX_t)=&\frac{1}{|\vm_t|}\sum_{i \in
\vm^{+}_t}[\NOC(i,\vX_t)-1],
\end{split}
\label{eq:match_cost}
\end{equation}
where $\vm^{-}_t, \vm^{+}_t \subset \vm_t$ denotes the sets of blobs
that have no corresponding blob and multiple corresponding blobs.
According to \Eq{eq:match_cost}, the configurations having most
blobs corresponding to one blob will have low cost. As shown in
\Fig{fig:consistency}, the configuration in (a) better accounts for
the blobs detected in each view, and it therefore has lower matching
costs.

\subsection{Kinetic coherency}
Motion tends to be smooth due to limited force agianst inertia.  This continuity
can be described by a kinetic model.

Consider a sequence of pairings that have already been assigned to
object $n$ up to time $t-1$,
$\vx^n_{1:t-1}=(\vx^n_1,\ldots,\vx^n_{t-1})$.  At time $t$,  a pairing
$\vp$ is to be  tentatively assigned to the object $n$.   To evaluate
this assignment $\vx^n_t = \vp$, we define a motion deviation
function $\fE^{\left<J\right>}_k(\vp)$ that measures the coherency
between the pairing $\vp$ and the $J$ previous pairings
$\vx^n_{t-J:t-1}$. If the deviation is small, the assignment is a
reasonable one with regard to previous assigned pairings.

 Taking all objects into account, the total kinetic coherency
at time $t$ of a given configuration sequence $\vX_{1:t} =
(\vX_1,\ldots,\vX_t)$ can be evaluated by
\begin{equation}\label{eq:kinetic_cost_old}
\fE^{\left<J\right>}_K(\vX_t) = \sum^N_{n=1}
\fE^{\left<J\right>}_k(\vx^n_t).
\end{equation}

At this time, we only consider the case that all objects are present
in the scene. In other words, no dummy pairing is assigned to any
object. Hence each pairing here corresponds to a 3D position or a
pair of 2D positions on image planes. In the following sections, we
present several deviation functions $\fE^{\left<J\right>}_k(\cdot)$
by using different kinds of kinetic models both in 3D space and on
2D image planes.
\subsubsection{Kinetic models in 3D space} If the intrinsic and extrinsic parameters are 
known for each camera, given a pairing, its 3D location can be
computed. Consider a sequence of pairings
$\vx^n_{1:t}=(\vx^n_{1},\ldots,\vx^n_t)$ continuously assigned to
object $n$. Denote their corresponding 3D locations by
$\vms_{1},\ldots,\vms_t$. We use some kinetic model to predict the
3D locations at time $t$ from $J$ previous locations, namely
$\hat{\vms}_t = \fpm(\vms_{t-J},\ldots,\vms_{t-1})$. Then the
deviation function is defined as
\begin{equation}
\fE^{\left<J\right>}_k(\vx^n_t) = \|\hat{\vms}_t - \vms_{t}\|.
\end{equation}
Selection of the kinetic model relies on priori knowledge of the
motion of the subjects. We adopt here two kinetic models that are
prevalently used in tracking moving objects.

The first one is nearest-neighbor model \cite{veenman2001rmc}. The
nearest-neighbor model depends on only one previous location and
uses this location as a prediction for current time, i.e., $J = 1,
\fpm(\vms_{t-1}) = \vms_{t-1}$.  The prediction error at time $t$
for the pairing sequence assigned to object $n$  is:
\begin{equation}
 \fE^{\left<1\right>}_k(\vx^n_t) = \|\vms_{t}-\vms_{t-1}\|.
 \label{eq:nearest_neighbor}
\end{equation}
The nearest-neighbor model has been proved to be effective in many
tracking applications, particularly when the motion is complex and
hard to be formulated, such as wandering people and drifting
insects.

The second kinetic model is the smooth-motion model
\cite{sethi1987finding}. It is based on the observation that
sometimes objects move smoothly due to their inertia. The smoothness
indicates their velocities tend to be constant in a local time span.
We compute the local velocity from locations at two previous time
instances and use it to predict the current location. That is, $J =
2, \fpm(\vms_{t-2},\vms_{t-1}) = 2\vms_{t-1} - \vms_{t-2}.$ The
deviation is then given by
\begin{equation}
 \fE^{\left<2\right>}_k(\vx^n_t) = \|2\vms_{t-1}-\vms_{t-2}-\vms_t\|.
 \label{eq:smooth_motion}
\end{equation}

\subsubsection{Kinetic models on 2D image planes}
The kinetic coherency can also be evaluated on image planes by
computing the prediction inaccuracies by 2D kinetic model in each
view. Consider a sequence of pairings
$\vx^n_{1:t}=(\vx^n_{1},\ldots,\vx^n_t)$ assigned to object $n$.
Denote the locations of their blobs in respective views by
$(\vms^{[1]}_1,\ldots,\vms^{[1]}_t)$ ,
$(\vms^{[2]}_1,\ldots,\vms^{[2]}_t)$. We define the deviation
function on 2D image planes based on nearest-neighbor model as
\begin{equation}
 \fE^{\left<1\right>}_k(\vx^n_t) = \sum_{v\in\{1,2\}}\|\vms^{[v]}_t-\vms^{[v]}_{t-1}\|.
 \label{eq:nearest_neighbor_2}
\end{equation}
Similarly, the deviation function derived from smooth-motion model
is defined as
\begin{equation}
 \fE^{\left<2\right>}_k(\vx^n_t) = \sum_{v\in\{1,2\}}\|2\vms^{[v]}_{t-1}-\vms^{[v]}_{t-2}-\vms^{[v]}_t\|
 \label{eq:smooth_motion_2}
\end{equation}

The advantage of using kinetic model on image planes is that it does
not require computing the 3D coordinates for each pairing. It
therefore enables our framework to be applied in weakly calibrated
cases where only fundamental matrices between cameras are known.

\subsubsection{Kinetic models at initial time steps}
Given a pairing sequence
$\vx^n_{t_0:t}=(\vx^n_{t_0},\ldots,\vx^n_t)$, the deviation can be
evaluated only when $t \geq t_0+J$ due to the definition of
$\fE^{\left<J\right>}_k(\cdot)$. We therefore use
$\fE^{\left<t-t_0\right>}_k(\cdot)$ to evaluate pairings at each
time step $t < t_0 + J$, and we let $\fE^{\left<0\right>}_k(\cdot) =
0$ at the initial time $t_0$.

Since we have prior knowledge that objects move in a way governed by
a given kinetic model,
 it is unnecessary to consider pairing sequences with large deviations from the model.
By setting a threshold, $\tsm$, if the motion deviation is larger
than $\tsm$, we let $\fE^{\left<J\right>}_k(\vx^n_t) = \infty$ and
discard this assignment.

\subsection{Visibility switching}\label{sec:visbility_switching}
Since an object may randomly enter and leave the scene,  their visibility state could change during the capture
process. We use a dummy pairing $\vnu$ to represent the absent state
of an object. So the problem of changing visibility state can be
handled by alternately assigning dummy pairing or non-dummy pairing
to the object.

When a dummy pairing and a non-dummy pairing are assigned to the
same object at successive time steps, visibility switching happens
on this object. We penalize visibility switching by introducing a cost
$\eta$ for each object. To encourage extracting continuous motion
trajectories, the value of $\eta$ is set larger than the motion
deviation threshold $\tsm$ described previously.

Taking the visibility switching into consideration, we evaluate the
total kinetic coherency of a configuration sequence $\vX_{1:t}$ at
time $t$ by
\begin{equation}\label{eq:kinetic_cost}
\fE^{\left<J\right>}_K(\vX_t) = \frac{1}{|N^*_{t-1}\cup
N^*_t|}\left[ \sum_{n \in
U_t}\fE^{\left<J\right>}_k(\vx^n_t)+\eta|V_t|\right]
\end{equation}
instead of \Eq{eq:kinetic_cost_old}, where $N^*_{t-1},N^*_t \in
\{1,\ldots,N\}$ denotes objects assigned with non-dummy pairings at
the time steps $t-1$ and $t$. The active objects at current time
step, $N^*_{t-1}\cup N^*_t$, consist of two types of objects. The
first type of objects, denoted by $U_t =(N^*_{t-1} \cap N^*_t)$, are
the objects now present in the scene without changing visibility
state at current time step.  The second types of objects, denoted by
$V_t = (N^*_{t-1}\cup N^*_t) \setminus U_t$, refer to the objects
now switching the visibility state.

From \Eq{eq:kinetic_cost}, we can see that visibility switching is
expected to happen only when no pairings can be assigned to the
object with motion deviation less than $\tsm$. Otherwise, keeping
the visibility state unchanged and selecting pairings with small
motion deviations are more desired.

\section{Cost function  and optimization method}\label{section:energy}
Additively combining the above-discussed three cues, we obtain the
overall cost function:
\begin{equation}\label{eq:cost_function}
\begin{split}
&\fE(\vX_{1:T})=
\\&\sum_{t=1}^T \left[\alpha\fE_E(\vX_t)+\beta_1
\fE_{C_1}(\vX_t)+ \beta_2 \fE_{C_1}(\vX_t)\right]+\gamma
\sum_{t=2}^{T}\fE^{\left<J\right>}_K(\vX_t)
\end{split}
\end{equation}
where $\alpha,\beta_1,\beta_2,\gamma$ are the parameters to control
the weights of these terms.

Obtaining the globally optimum result with respect to this cost function requires
evaluating all possible configuration sequences in the solution
space. Apparently optimization via brute-force enumeration is
computationally intractable due to the extremely high dimension of the
solution space. Noticing that the cost function can be recursively
decomposed into
\begin{equation}\label{eq:recursive_min}
\fE(\vX_{1:\,t}) =\fE(\vX_{1:\,t-1})+\Delta\fE(\vX_t),
\end{equation}
where the increment of cost $\Delta\fE(\vX_t)$ is
\begin{equation}
\Delta\fE(\vX_t) = \alpha\fE_E(\vX_t)+
\beta_1\fE_{C_1}(\vX_t)+\beta_2\fE_{C_2}(\vX_t) +
 \gamma \fE^{\left<J\right>}_K(\vX_t),
 \label{eq:incremental_cost}
\end{equation}
we are motivated that the cost function can be optimized
sequentially through cost propagation. We present an optimization
approach base on configuration proposition and cost propagation.
 It is accomplished iteratively at each time instance by two successive stages: 1) proposing possible configurations for
 next frame, 2) propagating
cumulative cost from previous ones to each newly proposed
configuration.

Since the number of possible configurations increases exponentially
with the number of objects, it is impossible to list all possible
configurations at each time step by brute-force enumeration. It is
also unnecessary to enumerate all configurations, because many of
them will lead to very high overall cost. Keeping only the
configurations with low costs at each time step will tremendously
decrease the overall computational cost.

In \cite{zou2010reconstructing}, Gibbs sampling \cite{geman1984srg} is
used to obtain these configurations with low costs by converting the
cost function into probability distribution function. But the
probability distribution tends to be sparse  and contains multiple
modes (local peaks),  the process is often stacked in a local mode
and waste lots of time in sampling configurations of no interest.
Instead of using Gibbs sampling, we use a method based on assignment
ranking to generate configurations with low costs more efficiently.
It is described in the following sections.

\subsection{Sampling configurations at initial frame}
At the first frame, we aim to obtain configurations with the
smallest costs according to the cost function $\fE(\vX_1)$. We
denote this sampling process by
\begin{equation}
\vX_{1}^{(k)} \sim \fE(\vX_1),
\end{equation}
where $k = 1,\ldots, K$ and $K$ is the number of configurations we
wish to sample. A configuration $\vX_1^{(k)}$ is a combination of
$N$ pairings chosen from $\vm^{[1]}_1 \times \vm^{[2]}_1 \cup
{\vnu}$. It is obviously impossible to evaluate all these
combinations to acquire the $K$-best ones in a reasonable time when
the number $N$ is large.

We know that a blob tends to have
one corresponding blob in the other view (\Sec{sec:one_to_one_match}). We convert this cue into
a one-to-one constraint and impose this constraint on choosing
pairings to generate configurations, then
 the sampling problem becomes obtaining $K$-best assignments
between the left blobs and the right blobs with the lowest costs in
\Eq{eq:epipolar_cost}. This may cause loses of some possible
configurations (some blob may happen to correspond to several blobs
in the other view because of occlusion), but it could tremendously
reduce the computational costs by avoiding enumerating possible
solutions in the entire configuration set.

 Let $a_{ij} \in \{0,1\} $ be an indicator of
the left blob $i$ being assigned to the right blob $j$ ($1$ for
true, $0$ for false). Denote the pairing composed of these blobs by
$\vp$. The cost of blob $i$ being assigned to blob $j$ is given by
$c_{ij} = \fG(\vp)$. Denote the number of blobs in each view by
$S_1,S_2$.  Without loss of generality, here we assume that $S_1
\leq S_2$.  Thus an assignment $\mathbf{a} = (a_{ij})$ should
satisfy the following conditions
\begin{equation}
\begin{split}
\sum^{S_2}_{j=1} a_{ij} = 1, \, i = 1,\ldots,S_1\, \text{and}\,
\sum^{S_1}_{i=1} a_{ij} \leq 1, \, j = 1,\ldots, S_2 \\
a_{ij} \geq 0.
\end{split}
\end{equation}
 Our goal is to obtain the $K$-best assignments
that have the lowest costs of
\begin{equation}
 \sum_{i}\sum_{j} a_{ij}c_{ij}.
\end{equation}
This is an assignment ranking problem that can be solved by Murty's
algorithm \cite{murty1968algorithm}, of which the computational
complexity is $O(K\cdot N^4)$, $N$ here is $S_1$. Using the improved
algorithms described in
\cite{pascoal2003note}\cite{pedersen2008algorithm}, the
computational complexity could reduce to $O(K\cdot N^3)$.

By solving the assignment ranking problem, $K$ configurations of
high interests can be acquired. For each sampled configuration , we
use \Eq{eq:cost_function} to compute its cost. Tree-like data
structure is used to store sampled configurations. The
configurations sampled at the initial frame are stored at the roots
of these trees. The trees grow when configurations at the next frame
are sampled from the parent configuration frame by frame.  In the
next stage, we described how to iteratively sample the subsequent
configurations of the current configuration and evaluate their
cumulative costs.

\subsection{Sampling child configurations at the next frame}
For each configuration at current time step $t-1$, we can obtain a
sequence of configurations by traveling back to the root of the
tree. Denote the configuration sequence by $\vX'_{1:t-1}$. We
attempt to expand the tree by sampling an configuration of low
incremental cost, namely,
\begin{equation}
\vX^{(k)}_{t} \sim \Delta\fE(\vX_{t}). \quad(\vX_{1:t-1}=
\vX'_{1:t-1}).
\end{equation}
Although the incremental cost relies all three cues as shown in
\Eq{eq:incremental_cost}, we achieve the goal of sampling
configurations of high interests in a greedy way : we first try to
obtain configurations that having low costs of eipolar constraint
and kinetic coherency; we then adjust newly obtained configurations
by enforcing each of them to satisfy the one-to-one match criterion
to reduce the incremental cost.

As we know in \Sec{sec:visbility_switching}, keeping visibility
unchanged will lead to low kinetic cost defined in
\Eq{eq:kinetic_cost}. The first step is to assign the objects
assigned with non-dummy pairings in $\vX'_{t-1}$ to the possible
true pairings with infinity epipolar costs \Eq{eq:epi_cost_single}.
 The cost of assigning $i$-th object to $j$-th pairing is given by
\begin{equation}
c_{ij} = \alpha \fE_e(\vx^i_t)+\gamma
fE^{\left<J\right>}_k(\vx^i_t)\quad (\vx^i_{1:t-1} =
\vx'^i_{1:t-1}).
\end{equation} Here $\vx^i_t$ represents the $j$-th pairing assigned
to the $i$-th object. This is also an assignment problem as stated
in the pervious section. We also obtain the $K$-best assignments by
using assignment ranking algorithms.

Sometimes in some acquired assignments,  the pairing assigned to an
object could lead to a large kinetic deviation greater than a preset
threshold, namely, $\fE^{\left<J\right>}_k(\vx^n_t) > \epsilon_k$.
In this situation, the object is much likely to be absent from the
scene at that time, so we replace the assigned pairing of this
object with the dummy pairing $\vnu$ instead. The replacement is
reasonable since it could reduce the kinetic cost.  From these
assignments, $K$ configurations can be constructed by filling each
dimension related to these objects with assigned pairings.

The second step is to recompose the newly acquired configurations so
as to make the resulting configurations satisfy the constraint of
one-to-one match. Let ${\vm^{[1]}}^{-}$ and ${\vm^{[2]}}^{-}$ be the
sets of blobs corresponding to no object in respective views. To
decrease the costs derived from the one-to-one match cue, we try to
assign pairings int the set of ${\vm^{[1]}}^- \times {\vm^{[2]}}^+$
to the unassigned objects in $N^- = N \setminus N^*$. The task can
be done exactly in the same way as in sampling configurations at the
initial frame. By solving the assignment ranking problem, $K'$
combinations of pairings  with the smallest epipolar errors are
obtained. Each combination corresponds to a scheme of assigning
pairings in ${\vm^{[1]}}^{-} \times {\vm^{[2]}}^{-}$ to the
unassigned objects in $N^-$.  After that, the $K$ configurations
acquired in the previous step are recomposed to generate $K\times
K'$ new configurations, each of which has very low incremental costs
of \Eq{eq:incremental_cost}.

\subsection{Pruning and cost propagation}
The cumulative cost of a newly sampled configuration $\vX^{(k)}_t$,
denoted by $\fE^*(\vX^{(k)}_t)$, which is the cost of the
configuration sequence in the path from the tree root to the current
leaf, is recursively computed from its parent configuration,
$\vX^{(k')}_{t-1}$, namely, $\fE^*(\vX^{(k)}_t) =
\fE^*(\vX^{(k')}_{t-1}) + \Delta\fE(\vX^{(k)}_t)$. The cost of the
tree root at the first frame is directly evaluated by using
\Eq{eq:cost_function}.

\begin{figure}[ht] \centering
\includegraphics[height=
0.23\textwidth]{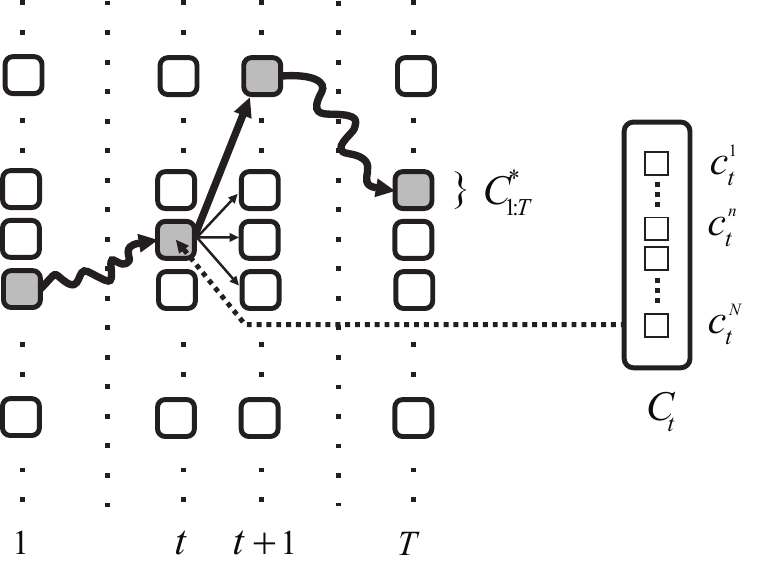}
  \caption{Optimization is done by sampling new configurations at next frame and
  propagating cumulative costs to them until reaching the last frame.
  Finally the optimum configurations sequence $\vX^*_{1:T}$ is obtained by tracing back from the configuration with the
  smallest cumulative cost.} \label{fig:joint_state_graph}
\end{figure}

As the number of configurations increase exponentially when the
trees grow over time, we prune the configurations with high
cumulative costs at each frame and keep the number of remaining
configurations in a reasonable level. Configuration sampling and
cost propagation are performed frame by frame  and finally the
optimum configurations can be acquired by tracing back from the
configuration with the lowest cumulative cost to the tree root at
the first frame, as shown in \Fig{fig:joint_state_graph}.

When the optimum configuration sequence $\vX^*_{1:T}$ is acquired,
3D motion trajectories can be computed from the pairing sequence in
each dimension of $\vX^*_{1:T}$. Notice that a pairing sequences
could contain some dummy pairings. These dummy pairings separate the
whole pairing sequence into several non-dummy pairing segments. The
motion trajectories are recovered at each of these segments for
corresponding objects.
\subsection{Parameters setting}
The variables $\alpha, \beta_1,\beta_2$, and $\gamma$ are four basic
parameters in our algorithm,  which control the weights of the four
components in the cost function. We define three variables
$\phi_e,\phi_{c},$ and $\phi_k$ to normalize the four components
into the same order of magnitude. Then we use four relative weights
$\tilde{\alpha} ,\tilde{\beta_1},\tilde{\beta_2},$ and
$\tilde{\gamma}$ that $\alpha = \tilde{\alpha}/\phi_e, \beta_1 =
\tilde{\beta_1}/\phi_c,\beta_2 = \tilde{\beta_2}/\phi_c,$ and
$\gamma = \tilde{\gamma}/\phi_k$ to exercise influence on the
respective costs. Here we set $\phi_e = \epsilon_e$ and $\phi_k =
\epsilon_k$. The value of $\phi_c$ depends on the density of objects
moving in the scene. It is set several times larger than the average
number of corresponding objects of each projection on the image
plane.

The threshold value of $\epsilon_e$ depends on the object detection
error and calibration error in the capturing system. Smaller
$\epsilon_e$ could lead to less possible pairings and therefore
reduce the running time for optimization, but it requires more
accurate object detection and calibration. The kinetic deviation
threshold $\epsilon_k$ can be determined in an automatic way. For
each pairing, we compute the distance to its nearest pairing at the
next frame using the distance functions between pairings as in
\Eq{eq:nearest_neighbor} or \Eq{eq:nearest_neighbor_2}. Denoting the
average nearest distance of all pairings by $ \bar{\mathscr{D}}$,
the threshold $\epsilon_k$ can be set empirically several times
larger than $\bar{\mathscr{D}}$, namely $\epsilon_k =
r_k\cdot\bar{\mathscr{D}}$, where $r_k > 1.0$.

The visibility switching cost $\eta$ is set to be larger than the
kinetic deviation threshold $\epsilon_k$ to prevent undesired
visibility switching, which usually breaks the continues
trajectories into pieces. The remaining parameters $K$ and
$\tilde{K}$ are the number of configurations sampled at each
sampling step and the maximum number of configurations retained at
each time instance. They are desirable to be large so as to produce
more favorable result, but not be exceedingly large because of the
limitation of the computational capability of hardware.  

\section{Experiments}
We have carried out experiments using simulated particle swarms and real-world fruit flies (Drosophila melanogaster). In the experiment 
of simulated swarms, the proposed method was compared with other strategies under different settings, including 
various density, velocity, and trajectory smoothness.  Since the ground truth was known, the performance was compared in a quantitative way.  In the fruit fly experiment, we recovered the 3D motion trajectories of a large group of 
flying fruit flies.  The performance was evaluated against visual inspection as the ground truth was unavailable. 

\subsection{Simulated particle swarms}\label{sec:simulated}
\begin{figure*}
\centering \subfigure[$\theta=0.0$]{\includegraphics[width =
0.15\textwidth]{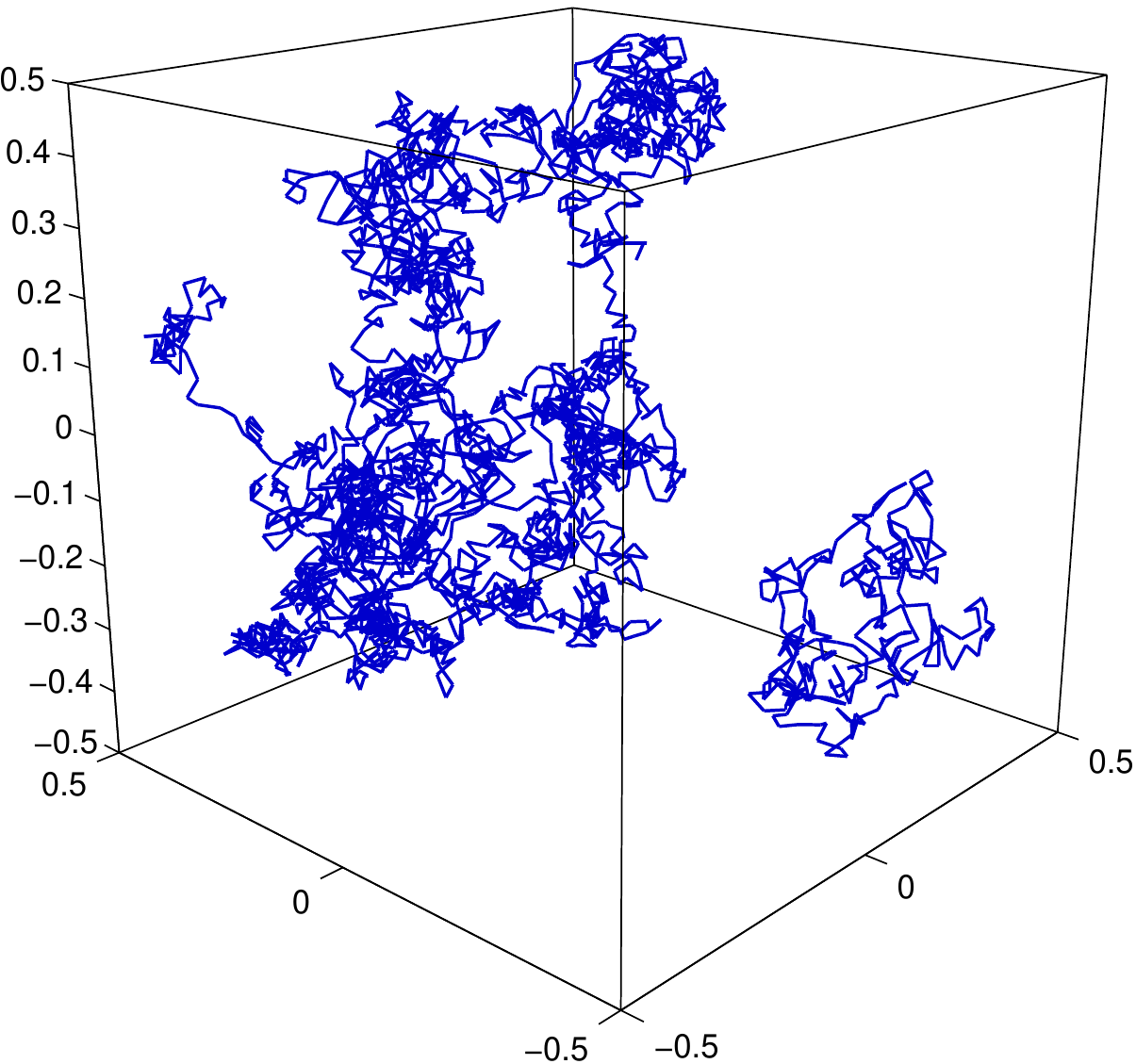}}\quad
\subfigure[$\theta=0.2$]{\includegraphics[width =
0.15\textwidth]{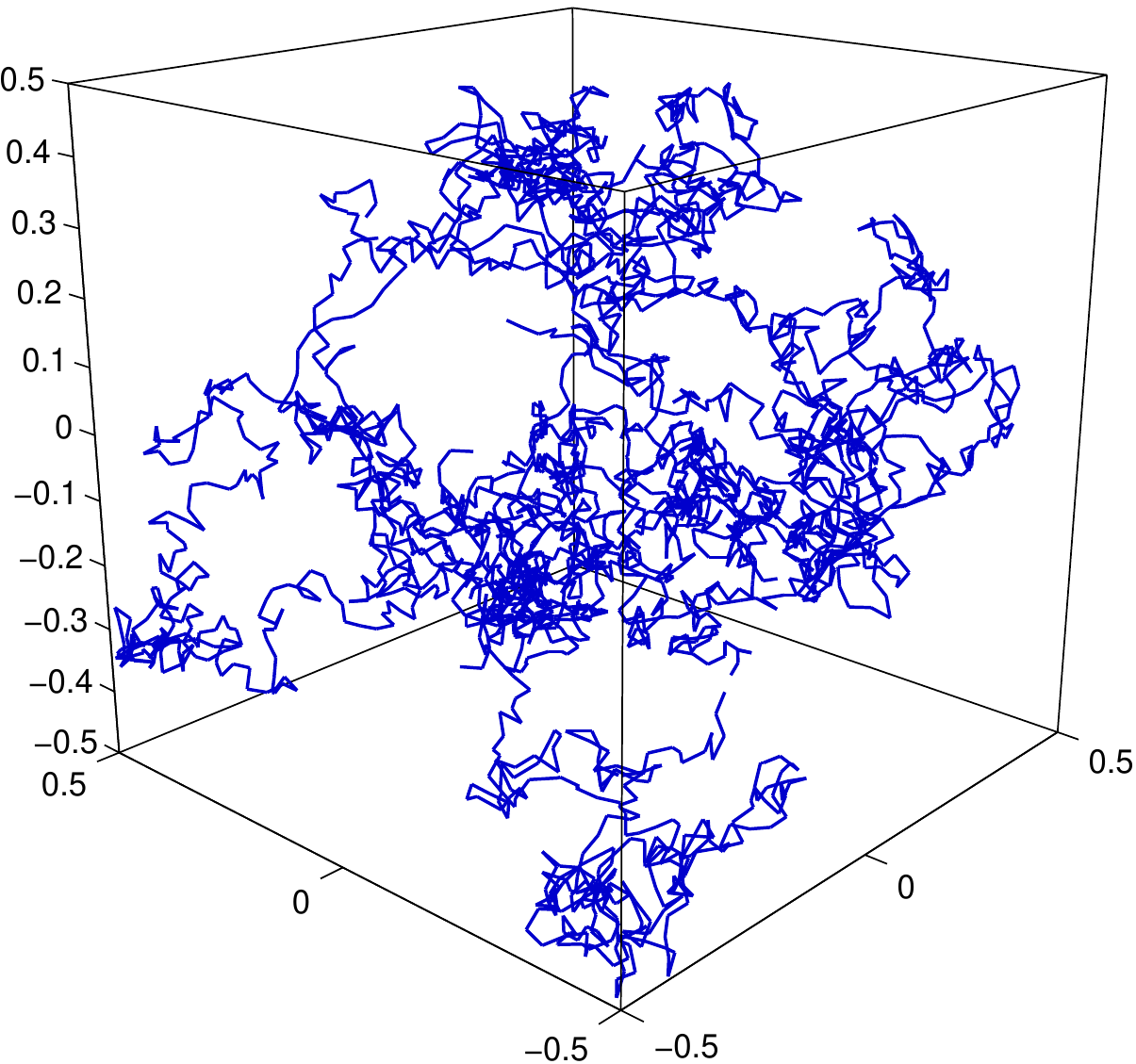}}\quad
\subfigure[$\theta=0.4$]{\includegraphics[width =
0.15\textwidth]{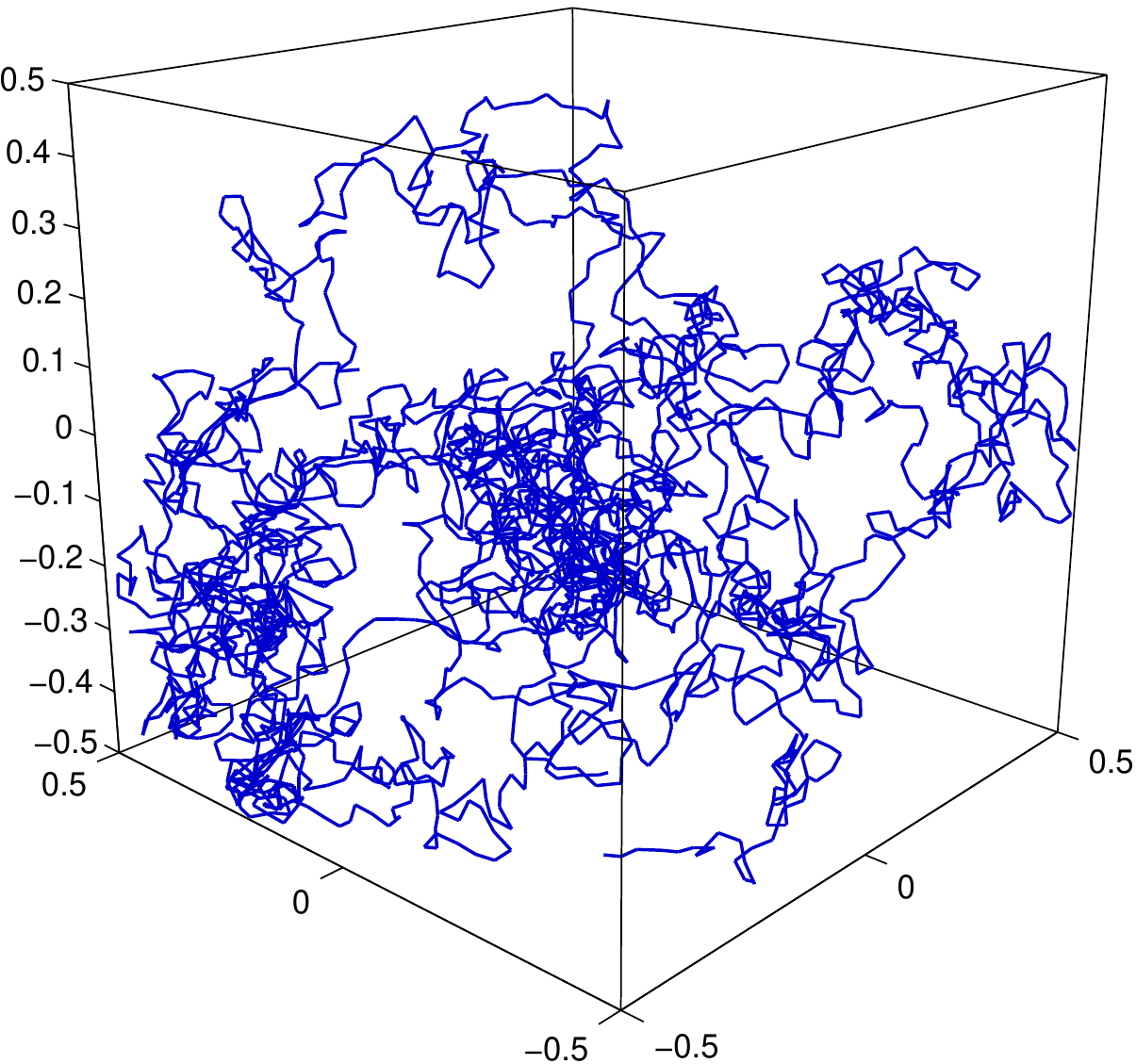}}\quad
\subfigure[$\theta=0.6$]{\includegraphics[width =
0.15\textwidth]{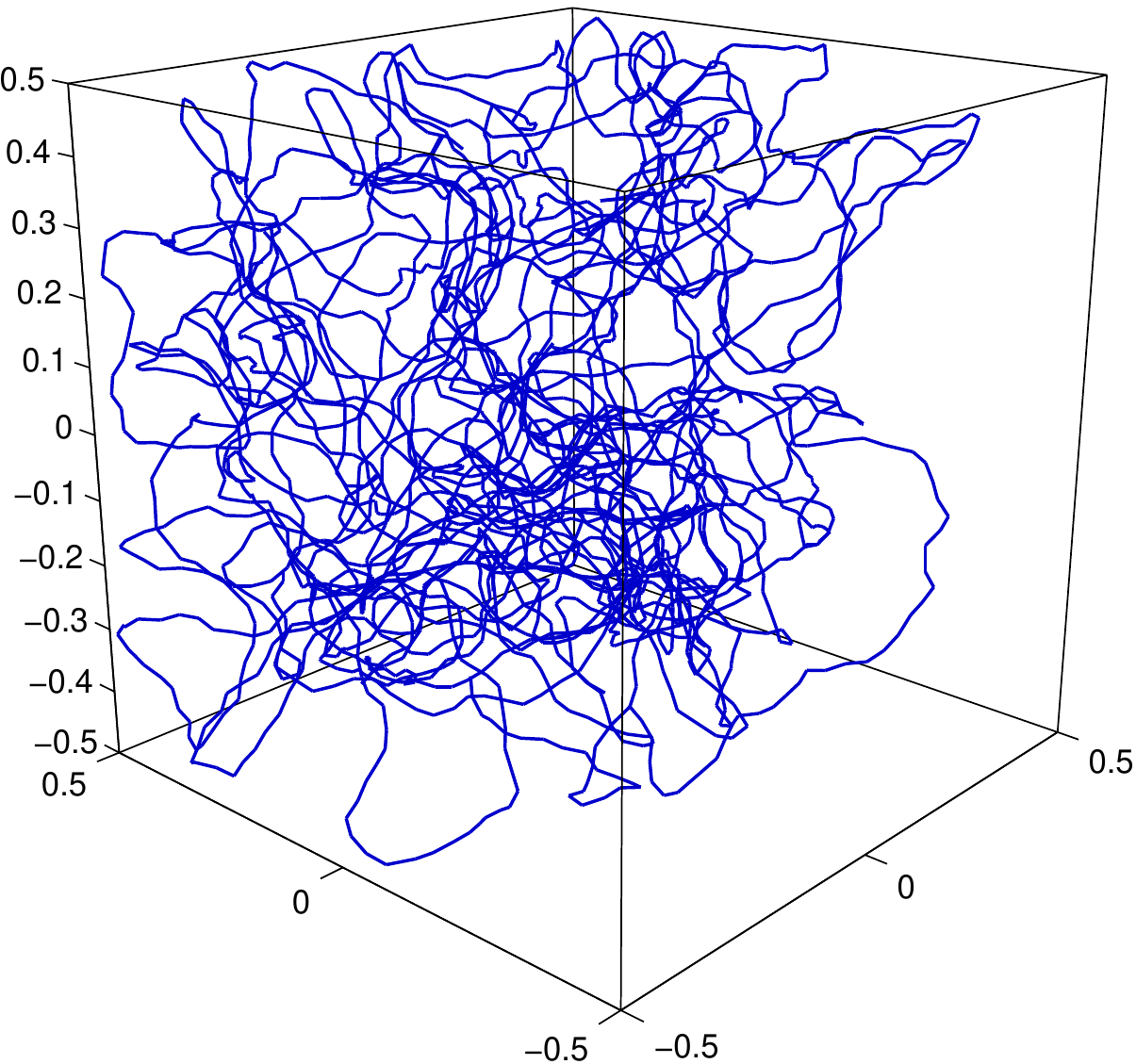}}\quad
\subfigure[$\theta=0.8$]{\includegraphics[width =
0.15\textwidth]{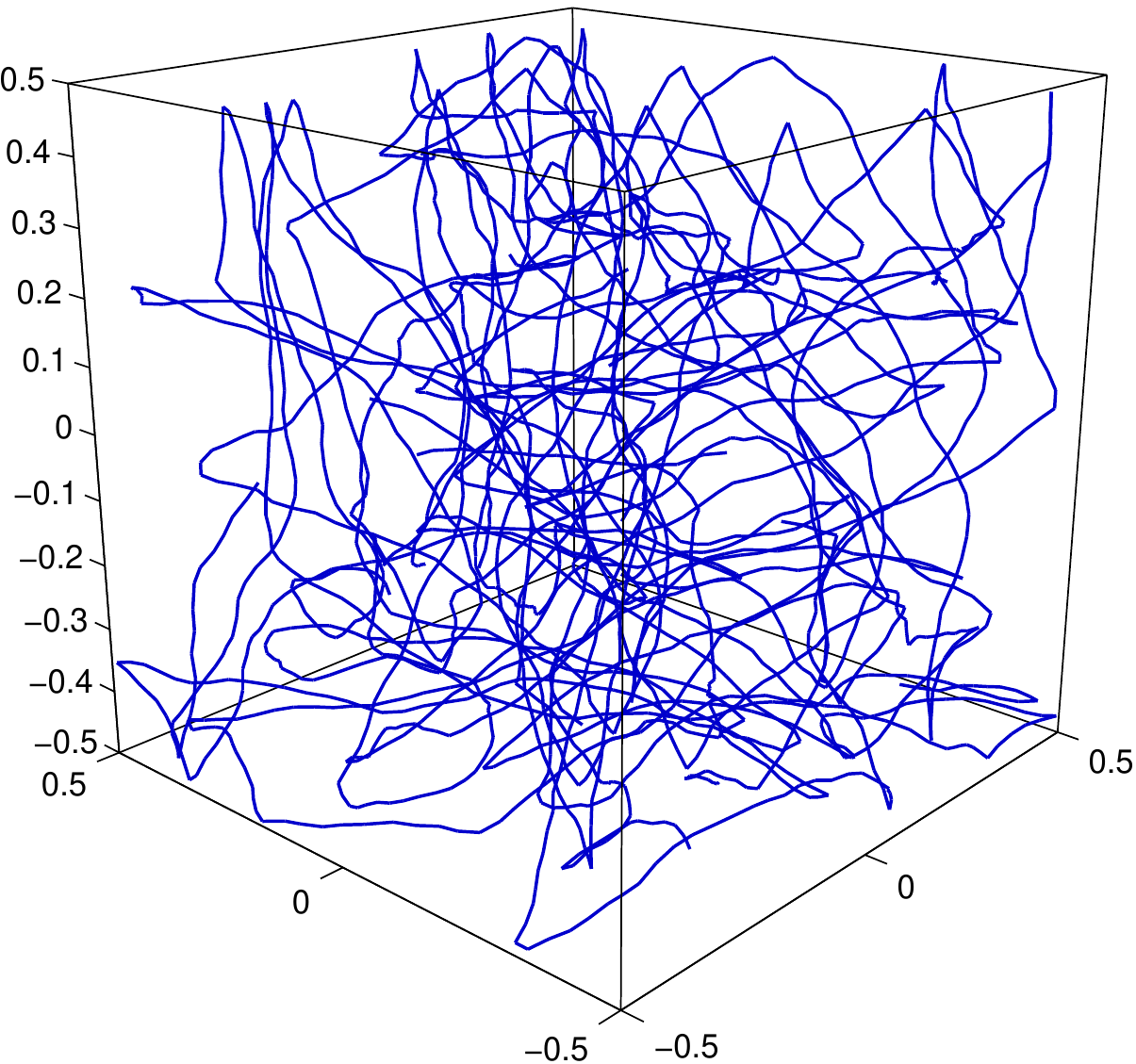}}\quad
\subfigure[$\theta=1.0$]{\includegraphics[width =
0.15\textwidth]{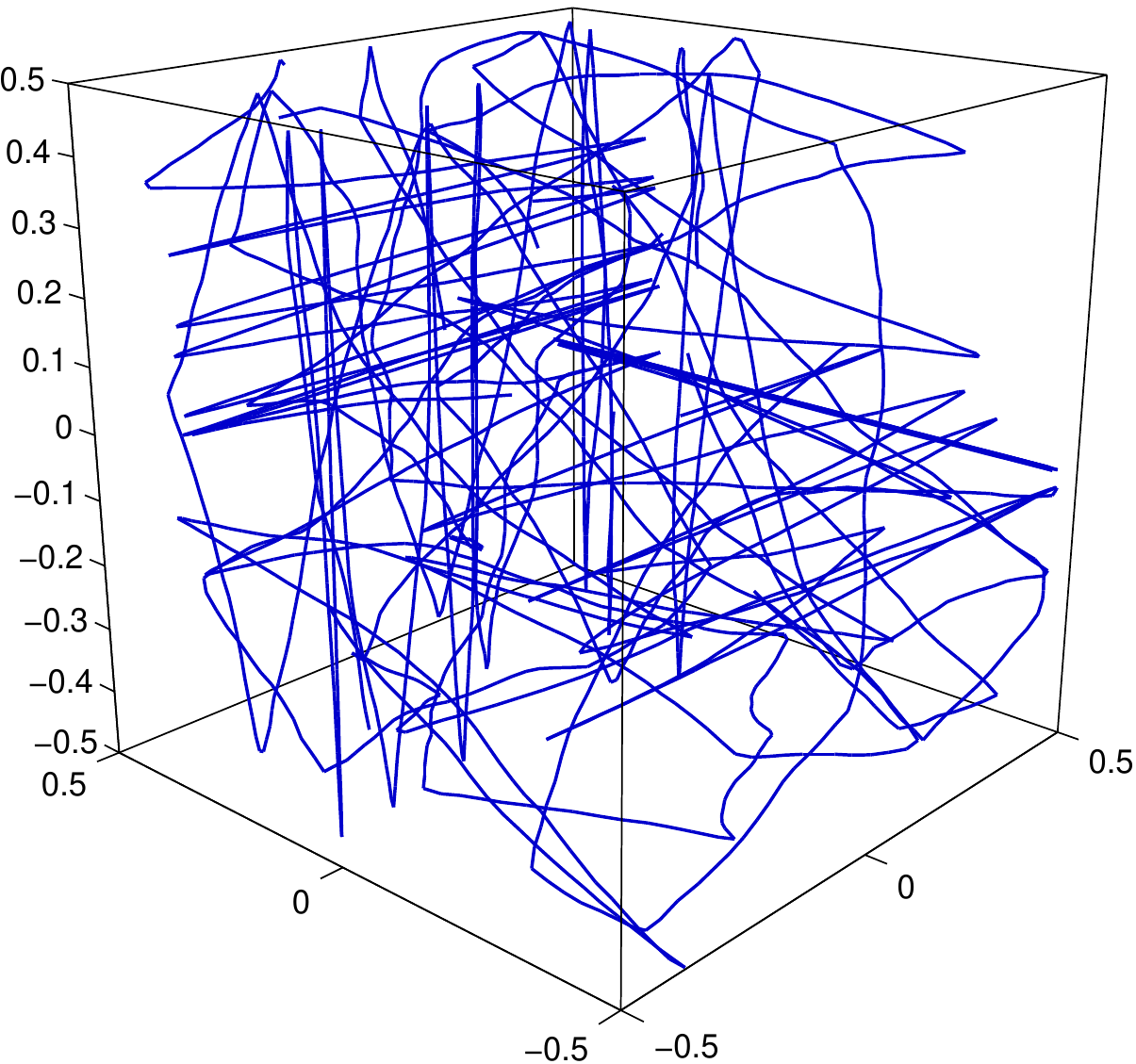}}\quad
 \caption{Trajectories generated with different $\theta$
}
 \label{fig:smoothness}
\end{figure*}

\emph{Data generation:} 
The particles are confined in  a
cube of size $1.0 m\times1.0 m\times1.0 m$. Each
particle is initialized with a random location and assigned  a
velocity of random direction and constant magnitude $v$.  At
each time step, the direction of the velocity vector is updated by
adding a random vector $\delta \sd$ followed by nomalization to make the vector of unit length,  that is
\begin{equation}
 \sd_t =  \frac{\theta \sd_{t-1} + (1-\theta) \delta \sd }{\|\theta \sd_{t-1} + (1-\theta) \delta \sd\|},
\end{equation}
The parameter $\theta \in [0,1]$ is used to control the smoothness
of the trajectory as shown in \Fig{fig:smoothness}. Each element
of $\delta \sd$ is uniformly drawn from $[-1.0,1.0]$. The resultant
velocity vector is then updated by
\begin{equation}
\vxv_t = (v_{t-1} + s_t) \sd_t.
\end{equation}
 $s_t \sim
where \mathcal{N}(0,0.05)$ is a Gaussian noise used to slightly perturb the
velocity magnitude.

After the velocity vector is obtained, the
location of the object is computed from previous location via $\vxx_t
= \vxx_{t-1}+\vxv_t \cdot \Delta t$. The objects keep on moving until they
hit the boundaries of the cube, where they bounce back.

The
scene is rendered using OpenGL frame by frame to produce video
sequences for the experiment.  
We use two virtual cameras of
resolution $800\times800$ to capture the scene.
Each test sequence contains $100$ video frames and each object is rendered as a sphere.  
The radius of each sphere is set to $0.005$ producing image blobs
of average diameter of $6$ pixels.

 We then treat the generated video sequences as input and  detect the blobs and compute their centers. To simulate detection
and calibration errors that exist in real-world cases, the obtained
coordinates are perturbed by random
noise drawn from a Gaussian distribution $s \sim \mathcal{N}(0,1.0)$
(pixel) (for both X and Y). 

\emph{Evaluation metrics}: The first step of evaluating 
performance is to match reconstructed trajectories
$\{r_1,\ldots,r_m\}$ to ground truth trajectories
$\{g_1,\ldots,g_n\}$. A reconstructed trajectory is regarded as well matched to a
ground truth trajectory  if the coordinates of the reconstructed
trajectory are nearly identical to those of the ground truth trajectory over the entire track.  Denote the position of $r_i$ at time $t$ by $r^t_i$ and the
position of $g_j$ at time $t$ by $g^t_j$. If $|r^t_i - g^t_j| <
\kappa$ for all time steps within the time duration of $r_i$, then
$r_i$ is said to be well matched to $g_j$ and the trajectory $r_i$ is said to be
\emph{correct}. The threshold $\kappa$ is used to allow for numerical inaccuracy.
If a correct trajectory has the same time duration
as that of the corresponding ground truth trajectory, it is also
\emph{complete}. The unmatched trajectories are false ones.

 The evaluation is performed by assessing
three aspects of the result : 1) the proportion of correct trajectories
(\emph{correctness}) ; 2)  among these correct trajectories, how
many of them are complete (\emph{completeness}); 3)
 the amount of false trajectories (\emph{precision}).

 Let $\mathcal{L}_g$ denote the total length of ground truth
 trajectories
and $\mathcal{L}_c$ represent the length of correct trajectories. A metric ($\ECT$) is defined to measure the level of \emph{correctness}, namely,
\begin{equation}
\ECT = \mathcal{L}_c/\mathcal{L}_g.
\end{equation}
A higher value of $\ECT$ indicates more correct trajectories are
obtained. $\ECT$ approaching to $1$ indicates all the
ground truth trajectories are correctly reconstructed. The \emph{
completeness} of the result is evaluated by
\begin{equation}
\ECP = n_c/n_g,
\end{equation}
where $n_c$ denotes the number of complete trajectories, $n_g$ is
the number of ground truth trajectories. It is insufficient to use
only $\ECT$ and $\ECP$ to evaluate the result, since the result with
high $\ECT$ and $\ECP$ could still be undesiable, i.e. , some of the reconstructed trajectories do not exist in the groud truth (false trajectories).  So the result is
further evaluated by the \emph{precision} metric ($\EPC$), defined
as
\begin{equation}
\EPC = \mathcal{L}_c/\mathcal{L}_r.
\end{equation}
$\EPC$ measures the proportion of the correct trajectories out of all
reconstructed trajectories.  A high \emph{precision} indicates that
few false trajectories are generated. Ideally, if all trajectories
are correctly reconstructed and no false trajectory exists,
all metrics reach up to the maximum value $1.0$.

\emph{Methods for comparison}: Although there is no existing method highly
effective for reconstructing 3D motion trajectories of particle
swarms,  we implement several of them using three typical strategies
to compare with the proposed method. They are
Reconstruction-Tracking method, Tracking-Reconstruction method, and
Tracking-Reconstruction-Stitching method.

 In Reconstruction-Tracking (RT) method,  3D coordinates are reconstructed at each frame and
the 3D motion trajectories of the objects are then obtained via tracking in
3D space. The stereo correspondence between image objects across
views is established by solving a linear assignment problem that
minimize overall epipolar error. For tracking in the 3D space,
the state-of-the-art methods for tracking point-like objects, GOA
tracker, is used.

In Tracking-Reconstruction method (TR), objects are tracked on the
image plane of each view;  the
corresponding 2D trajectories are then matched by examining if their
coordinates constinuously satisfy the epipolar constraint over the entire track \cite{du2007rem}. After that, 3D trajectories are
reconstructed from each pair of matched 2D trajectories. We
use GOA tracker to track these 2D particle-like objects.

The Tracking-Reconstruction-Stitching (TRS) method is an improvement of
the Tracking-Reconstruction method. In TRS, we do not attempt to
track objects throughout the entire duration since the presence of tracking
ambiguity could lead to incorrect tracking results. Instead, we
track objects only in a local time span when no tracking ambiguity
exists - the object under tracking keeps a considerable distance
away from others at that time.  Then stereo matching is proceeded on
these 2D trajectory segments obtained by partial tracking to yield
trajectory segments in 3D space. In the final stage, these 3D
trajectory segments are stitched together to generate complete
trajectories.

\begin{figure*}
\centering
\includegraphics[width = 0.8\textwidth]{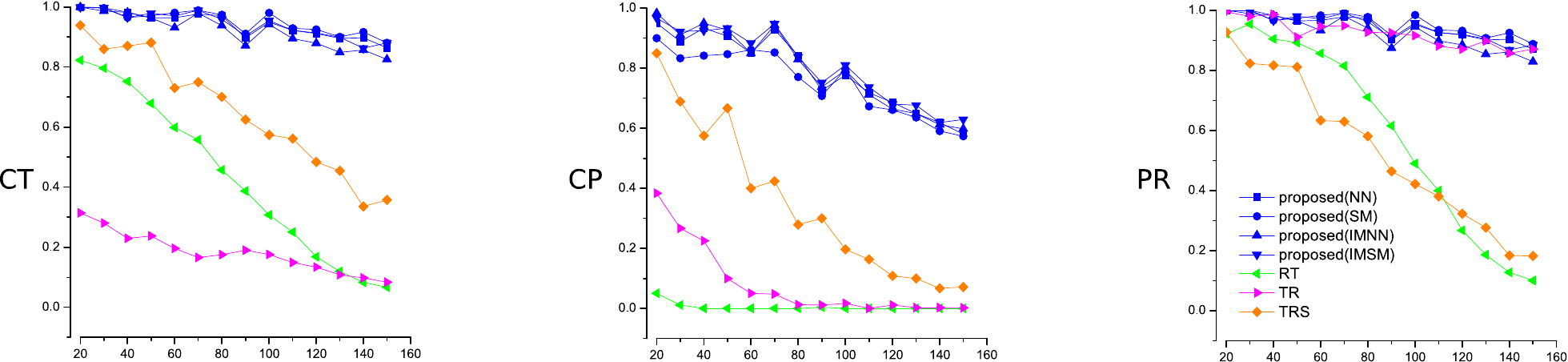}
\caption{Results for various object density. } \label{fig:res_diff_num}
\end{figure*}

\subsubsection{Experiment with various object density}
In this experiment, we analyze the performance of the proposed
method with respect to object density.  The speed magnitude is set to $v = 0.02$, and the
smoothness parameter is set to $\theta = 0.7$. The number of
objects present in the scene increases from $20$ to $150$ at a step of $10$.

We generate three data sets for each density and obtain the average
performance.
Four different kinetic models described in equations \Eq{eq:nearest_neighbor},
\Eq{eq:smooth_motion}, \Eq{eq:nearest_neighbor_2},
 and \Eq{eq:smooth_motion_2} (NN, SM, IMNN, IMSM for short respectively) are adopted for the
 test.
In the TR method, we use GOA tracker with smooth motion model to
track trajectories on the 2D image planes. The maximum number of
video frames allowed for stitching and interpolation for the TRS
method is set to $10$.

\begin{figure*}
 \centerline{
\subfigure[RT]{
\includegraphics[width = 0.23\textwidth]{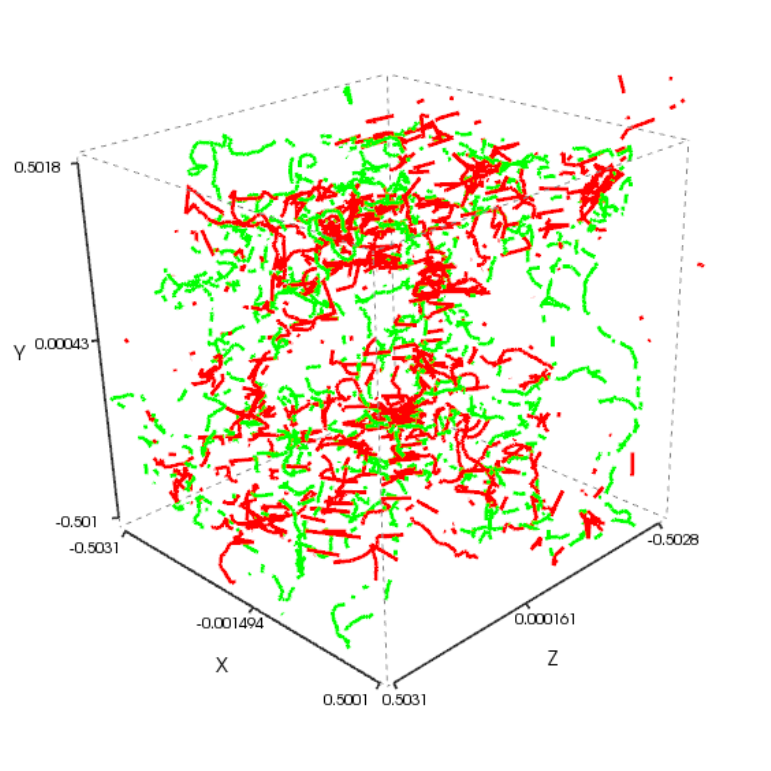}}
\subfigure[TR]{
\includegraphics[width = 0.23\textwidth]{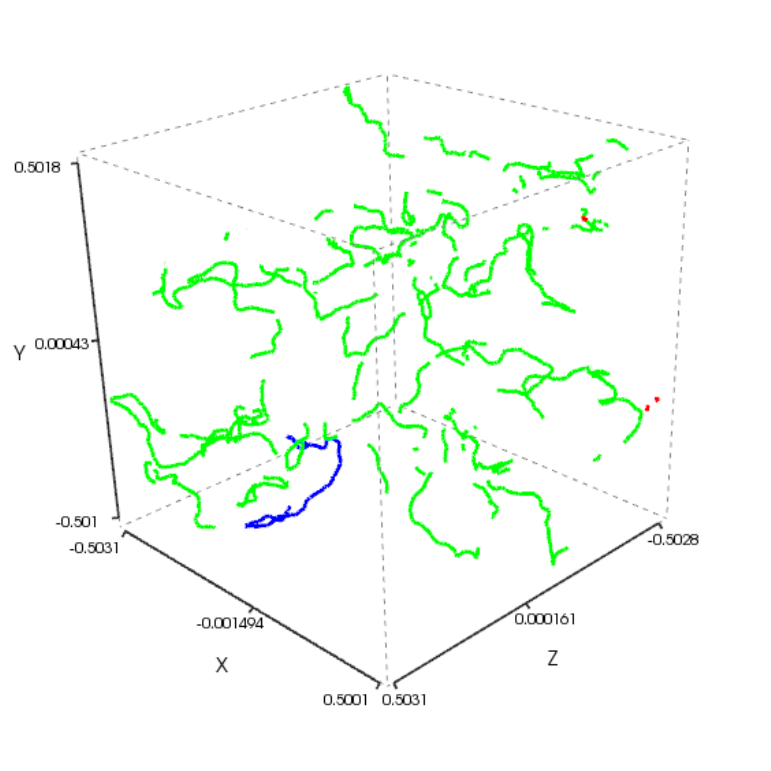}}
\subfigure[TRS]{
\includegraphics[width = 0.23\textwidth]{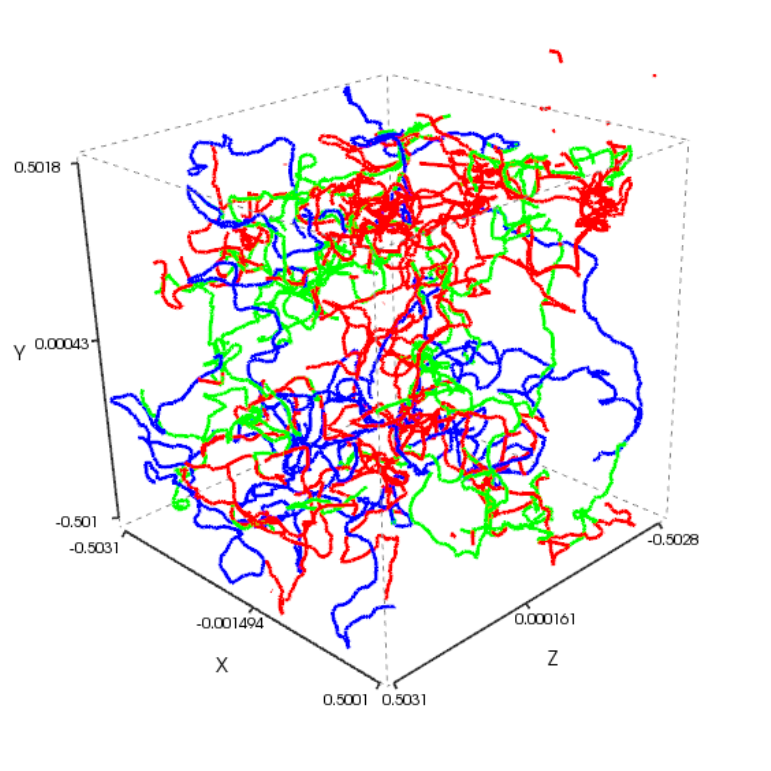}}
\subfigure[Proposed]{
\includegraphics[width =
0.23\textwidth]{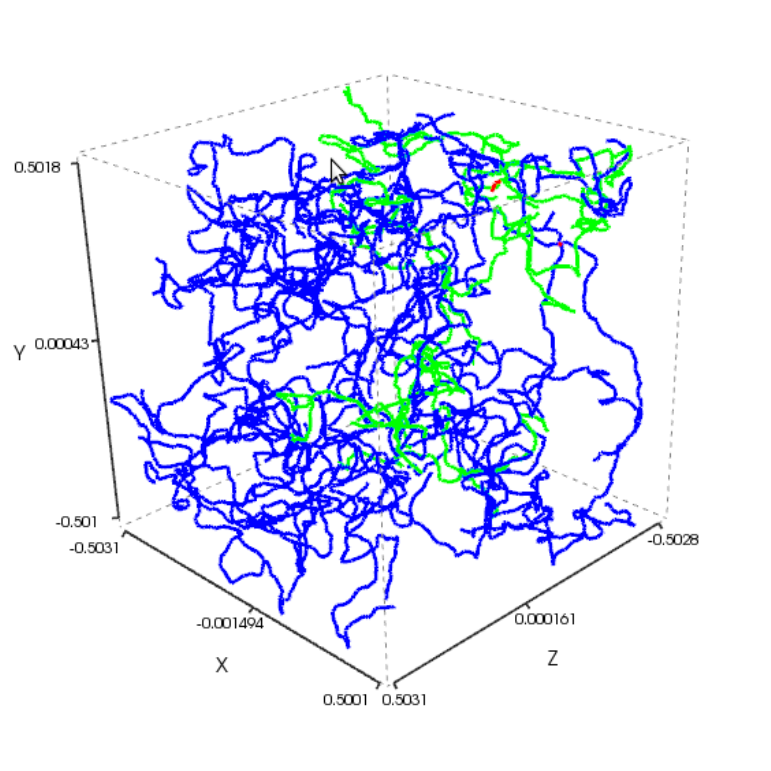}}} \caption{A
visual comparison of different methods ($\theta = 0.7$, $v = 0.02$,
$80$ objects). The blue ones show complete and correct trajectories; the green curves show \emph{correct}
trajectories; the blue ones represent the \emph{complete}
trajectories; the red ones are \emph{fake} trajectories.
}\label{fig:res_diff_methods}
\end{figure*}

The comparison results are shown in \Fig{fig:res_diff_num}.
Compared with the RT method, the TR method achieved a better $\ECP$, which
means it produces more complete trajectories than the RT method
does. The TR method however has a much lower $\ECT$. This is because the
TR method heavily relies on the 2D tracking performance.  Incorrect 2D tracks will cause matching failure and loss of
3D trajectories. The TR method nevertheless seldom generate false
trajectories ( see the result produced by the TR
method in \Fig{fig:res_diff_methods}). So the TR method gained a
high $\EPC$ as shown in \Fig{fig:res_diff_num}(c).

The TRS method performed better than the TR method and the RT method. The performance achieved by the TRS method declines quickly
when the density rises.  This is due to two reasons: 1) when
the density rises, the 2D track segments are too short to
eliminate stereo matching ambiguities; 2) it is difficult to connect
the broken trajectories because of the long time durations of the
missing pieces.

The results show that the proposed method performed best among
the compared methods. 
 When the number of
objects reached $150$,  $\ECP$ declined to $0.6$, while $\ECT$ and $\EPC$ of the result were still greater than $0.85$
.  The high $\ECT$ and $\EPC$ value
indicates that if some extrapolation processes are adopted, the
performance of the proposed method can be further improved - more
complete trajectories can be obtained. A visual comparison of
the various methods is also given in \Fig{fig:res_diff_methods}.

\begin{figure*}
\centering
\includegraphics[width = 0.8\textwidth]{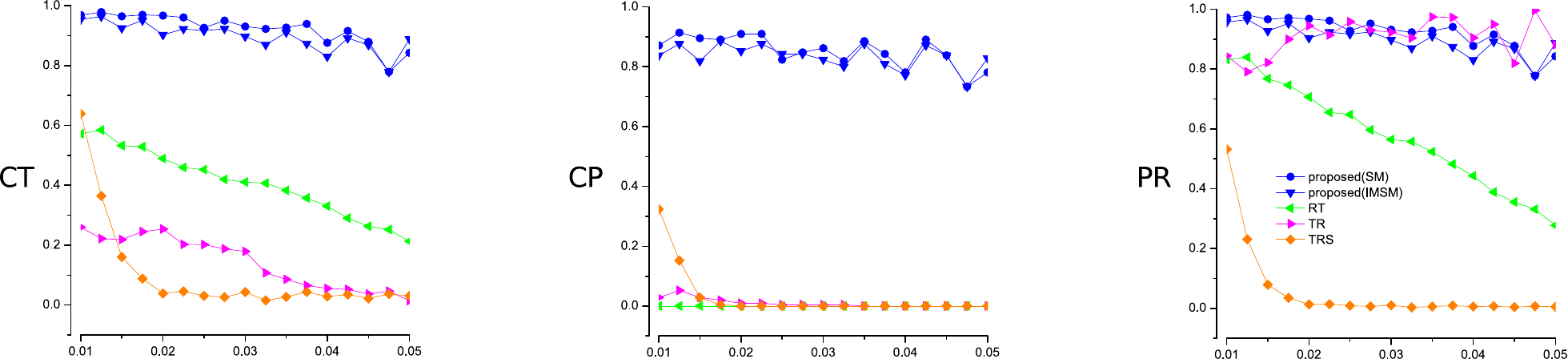}
\caption{Results for variable velocity. 
}
 \label{fig:res_diff_speed}
\end{figure*}

\subsubsection{Experiment with various velocity}
We change the mean velocity by modifying the velocity magnitude $v$
from $0.01$ per video frame to $0.05$ per video frame. The number of
objects in the scene is set to $70$. The smoothness parameter $\theta$
is set to $0.7$. At each step, the velocity magnitude increases
by $0.005$, and three data sets are generated to gauge the average
performance. Smooth motion models (SM, IMSM) are adopted for the
proposed method. The parameter settings of the methods for
comparison are the same as those in the previous experiment.

The results are shown in \Fig{fig:res_diff_speed}.  As the average
velocity of moving particles increases, the performance of the methods
for comparison drop dramatically. When the velocity rises greater
than $0.02$ (the particle travels two times its diameter between two adjacent frames ), no
complete trajectory could be generated by these methods.
The proposed method however  performed remarkably better than all
these methods. It still successfully reconstructed $80$ percent of
the ground truth trajectories and generates less than $20$ percent
of incomplete or incorrect trajectories when the velocity reaches to
$0.05$ per frame. This shows the strong capability of the proposed
method in handling video sequences of low frame rates with respect
to the velocity of the fast moving objects.

\subsubsection{Experiment with various smoothness}
 The trajectory smoothness is controlled by the parameter $\theta$ as discussed in \Sec{sec:simulated}. We set the number of
objects in the scene to $50$ and change the smoothness by adjusting
the parameter $\theta$ from $0.1$ to $1.0$ at a step of $0.1$.  Three data
sets are generated to gauge the average performance for each
$\theta$ value.

\begin{figure*}
\centering
\includegraphics[width = 0.8\textwidth]{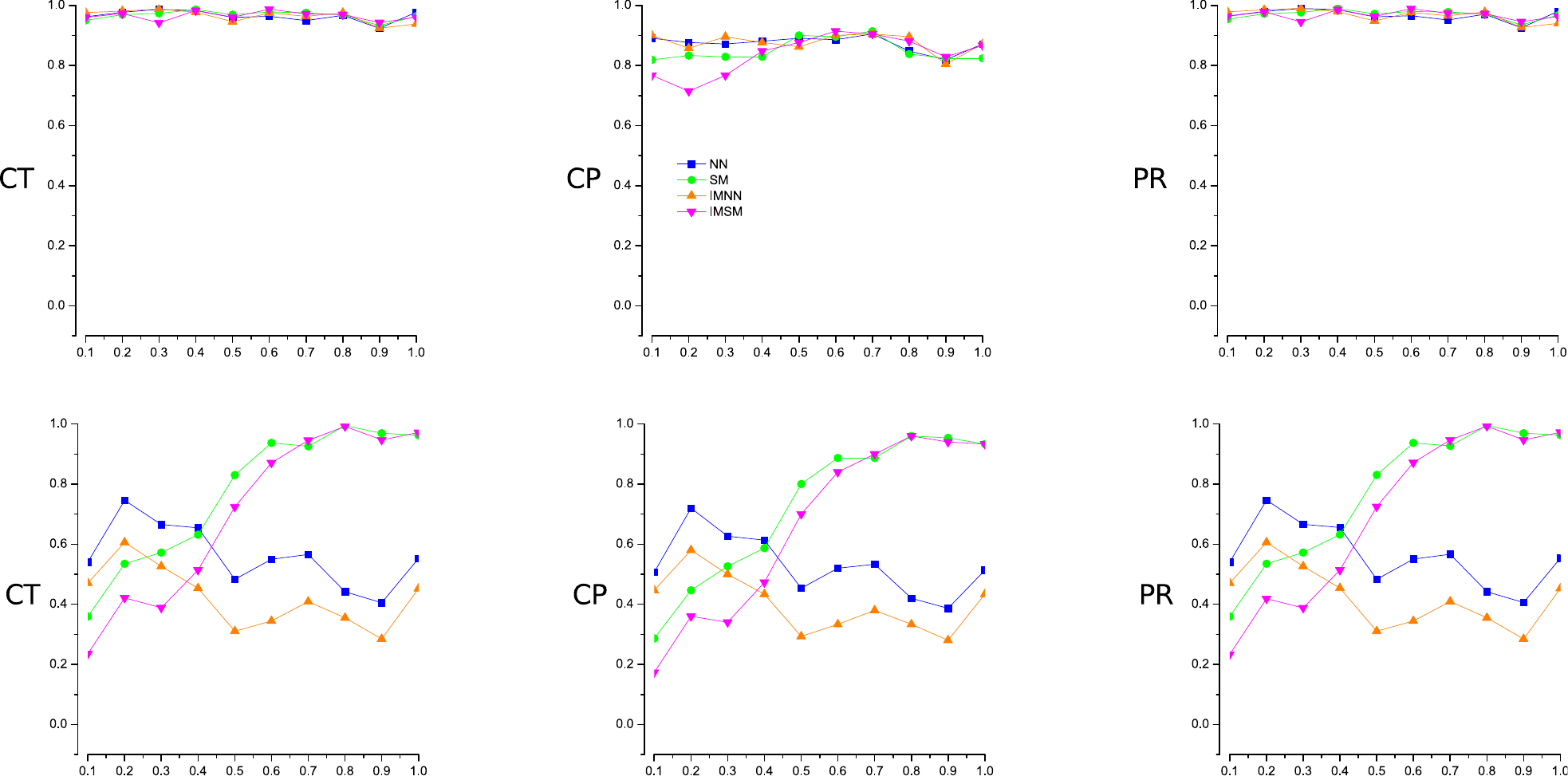}
\caption{
The results of different smoothness under two velocities - first row (0.01) and second row (
0.04). } \label{fig:res_diff_theta}
\end{figure*}

This experiment gives results under two velocity, $v = 0.01$,
$v=0.04$, as shown in \Fig{fig:res_diff_theta}. The results indicate
that when the average speed of moving objects is high,
selecting a proper kinetic model for reconstruction of trajectories
of specific smoothness
 plays an important role in gaining better performance. But in low speed cases,
the choice of kinetic models has less influence on the
performance. As shown in \Fig{fig:res_diff_theta}, when the average
velocity is $0.01$,  the method adopting two different
kinetic models produced similar performance.

\begin{figure}
 \centering
\includegraphics[width = 0.37\textwidth]{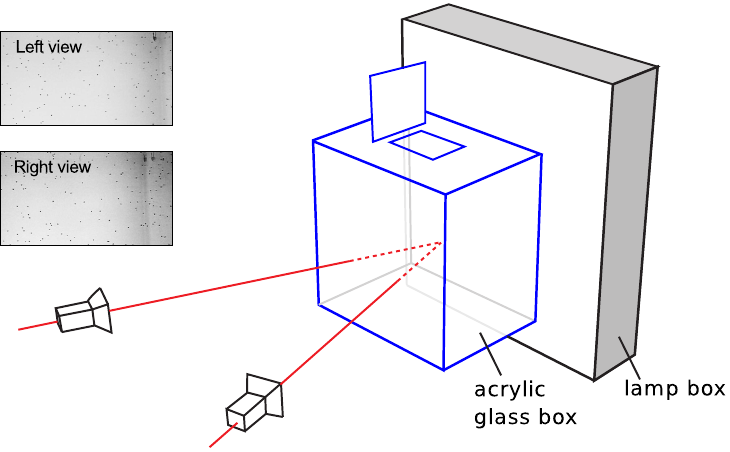}
\caption{The two camera system for fruit flies.}
\label{fig:system}
\end{figure}
\subsection{Fruit fly swarm}
Fruit fly (Drosophila Melanogaster) is a model organism extensively used in
genetic research since Thomas Hunt Morgan founded his famous Fly
Room in 1910\cite{kohler1994lords}.  It is also an important
subject for animal behavior research. Existing studies have been
conducted on the behavior of an individual fly or the
interaction of a small number of flies during sleep and
courtship\cite{vosshall2007into}, while the behavior in a large group
context has been little studied.

Although fruit fly is less well known for their social  interaction than classic eusocial 
insects such as bees and ants, their collective behavior has attracted increasing attention recently. A recent editorial 
\cite{naturemethods2009}, for example, has pointed out \emph{``methods to study the behavior of
Drosophila sp. in the context of a group may deepen our
understanding of the neural mechanisms underlying social behavior''}.

There are many inetersting questions 
waiting to be addressed: How do fruit flies interact among themselves? Is there any communication among them during flight? How do they avoid collisions when they are hovering
in a dense aggregation?  To answer such questions,  the ability to measure the 3D
motion trajectories of the individuals is of key importance.  This is therefore our motivation to reconstruct 3D motion
trajectories of a swarm of flying fruit flies.

Computer vision techniques had been used to acquire the 3D
trajectories of flying flies in \cite{tammero2002ivl,dickinson2008}.
They however mainly focused on the study of the flight performance
of an individual fruit fly and their systems captured only a few
flies. We set up a two camera system and applied the proposed approach to acquire
the 3D trajectories of hundreds of flying fruit flies.

Our experiment system for fruit flies is illustrated in \Fig{fig:system}. The fruit
flies flied in an acrylic glass box of size $35$cm $\times$ $25$cm
$\times$ $35$cm (height), where the background was uniformly
illuminated by a lamp box consisting of $6$ fluorescent tubes
and a frosted surface. To avoid flickering, these tubes were powered by a direct current supplier.
 Two Sony HVR-V1a
video cameras, working in the high speed mode of 200 frame per
second, were used to capture the scene from two different views. The
image resolution was $960 \times 540$. The two cameras were
temporally synchronized and geometrically calibrated.

 \begin{figure}
 \centering
 \includegraphics[width = 0.36\textwidth]{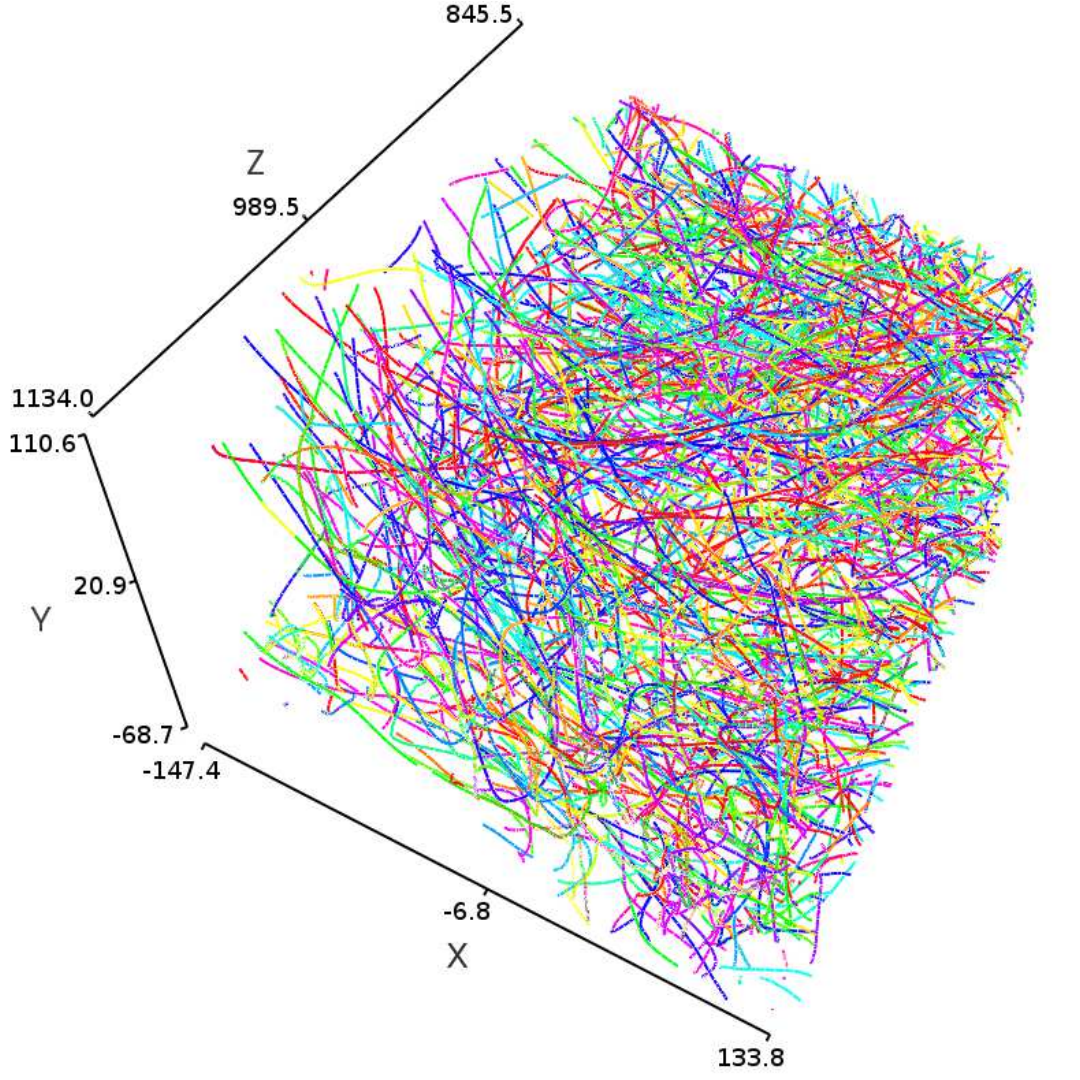}
 \caption{The reconstruction results from a pair of stereo video streams of $1000$ frames : $1105$ fly trajectories are reconstructed over a $5$ second period. The axis unit is $mm$.}
 \label{fig:all_trjs}
 \end{figure}

 \begin{figure}
\centering
\includegraphics[width = 0.36\textwidth]{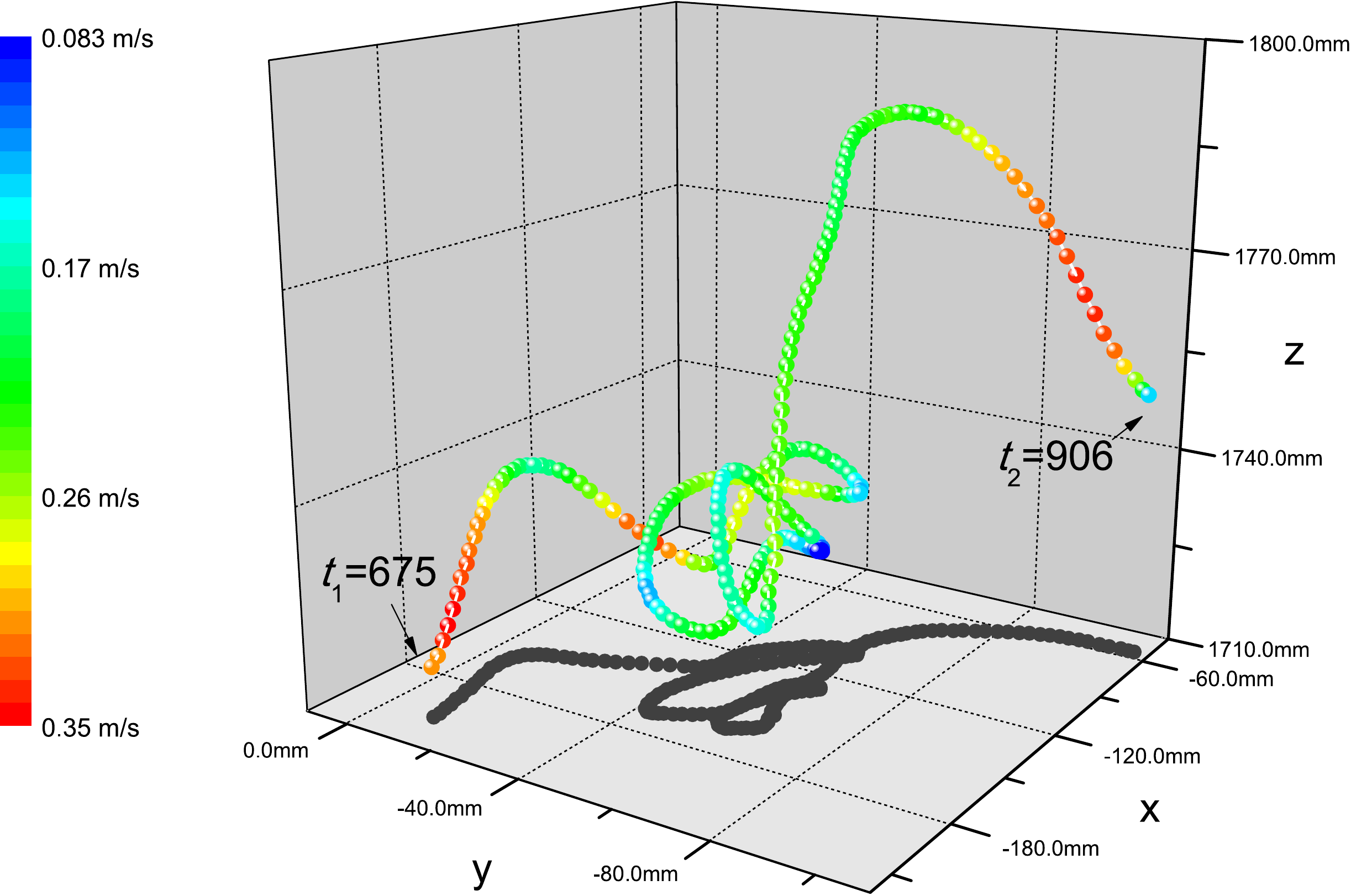}
\caption{The trajectory of a single fly entering the scene at the
$675$-th frame and leaving at the $906$-th frame.
Six saccades
\cite{fry2003aerodynamics} occured along the path. }
\label{fig:single_trj}
\end{figure}

 We applied background substraction technique to detect the fruit flies and then computed their centers.
We then applied our method to reconstruct the 3D trajectories of a
sequence of $1000$ frames ($5$ seconds). 
 The results are given in \Fig{fig:all_trjs}, showing $1105$ reconstructed
trajectories. 

\begin{figure*}
 \centering
  \begin{tabular}{c@{ }c@{ }c@{ }c@{ }c@{ }c@{ }c@{ }c@{ }c}
  \includegraphics[width=0.105\textwidth]{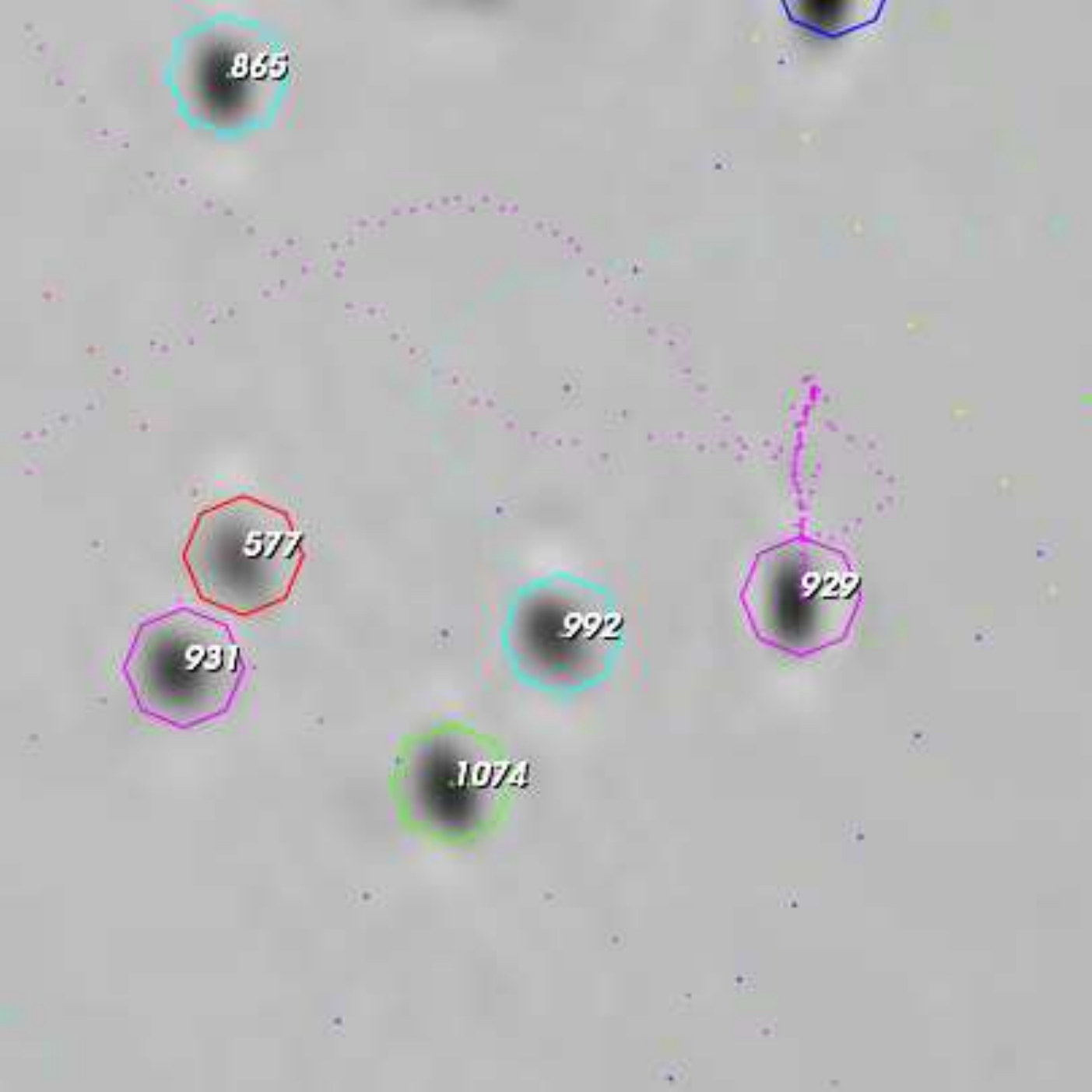}
 \includegraphics[width=0.105\textwidth]{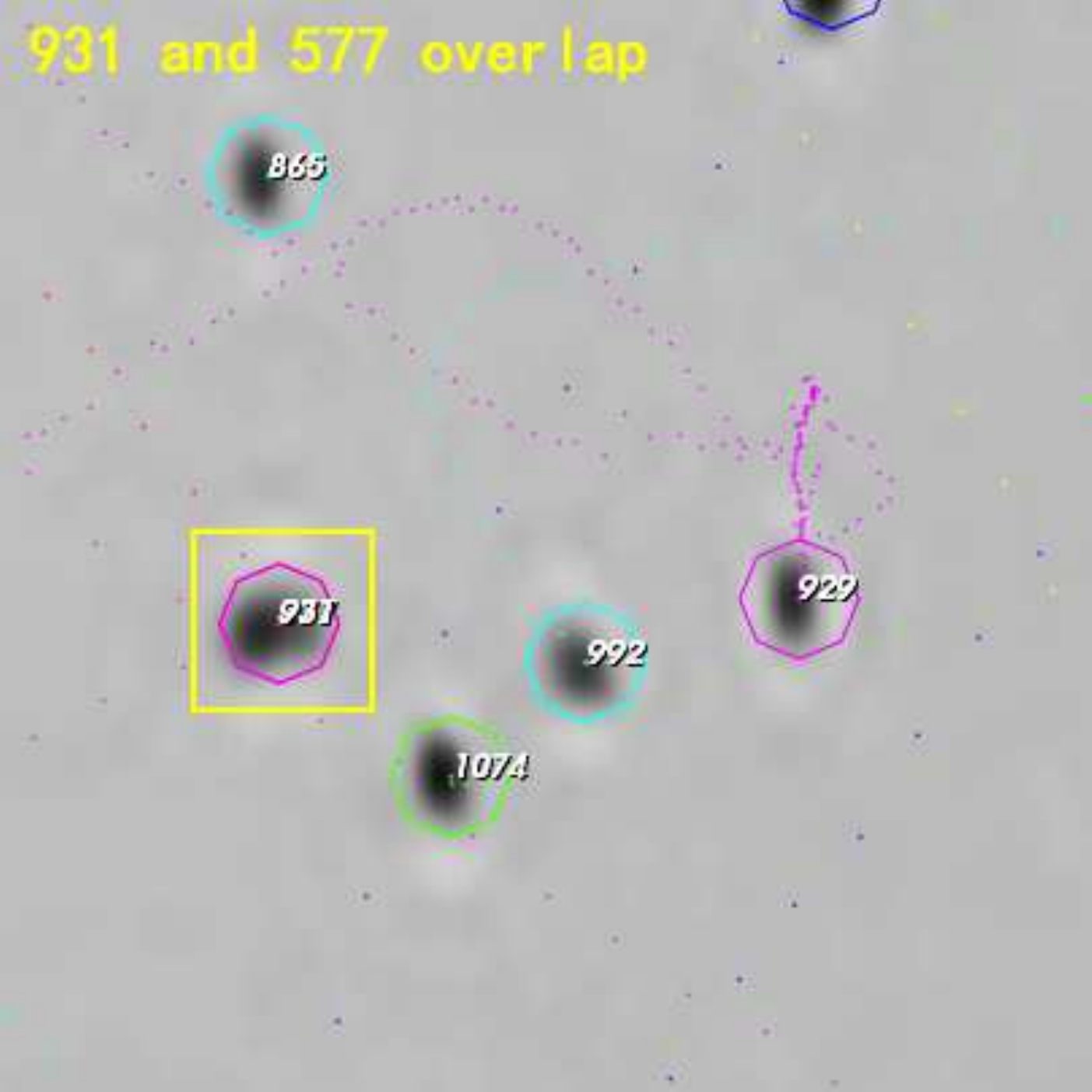}
 \includegraphics[width=0.105\textwidth]{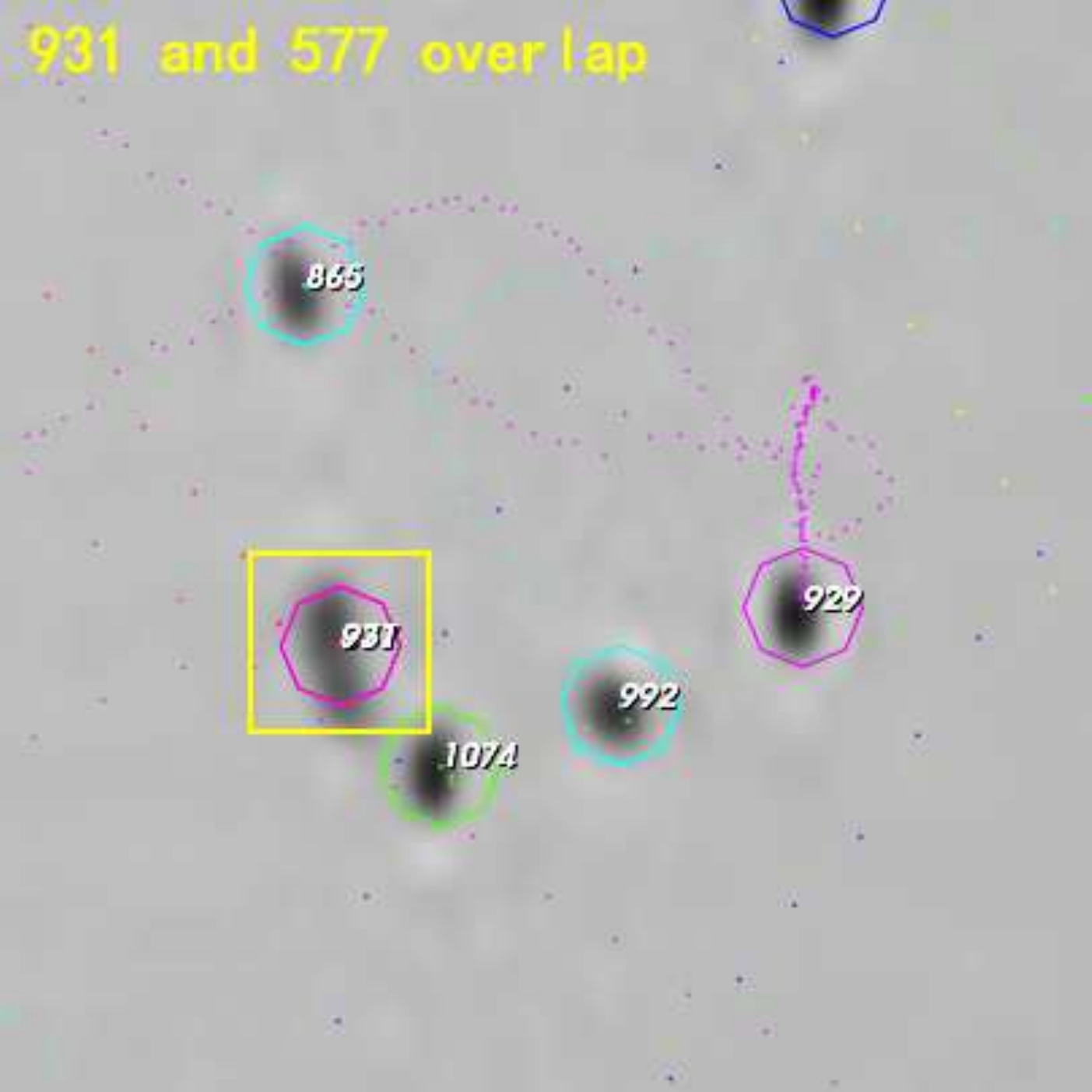}
 \includegraphics[width=0.105\textwidth]{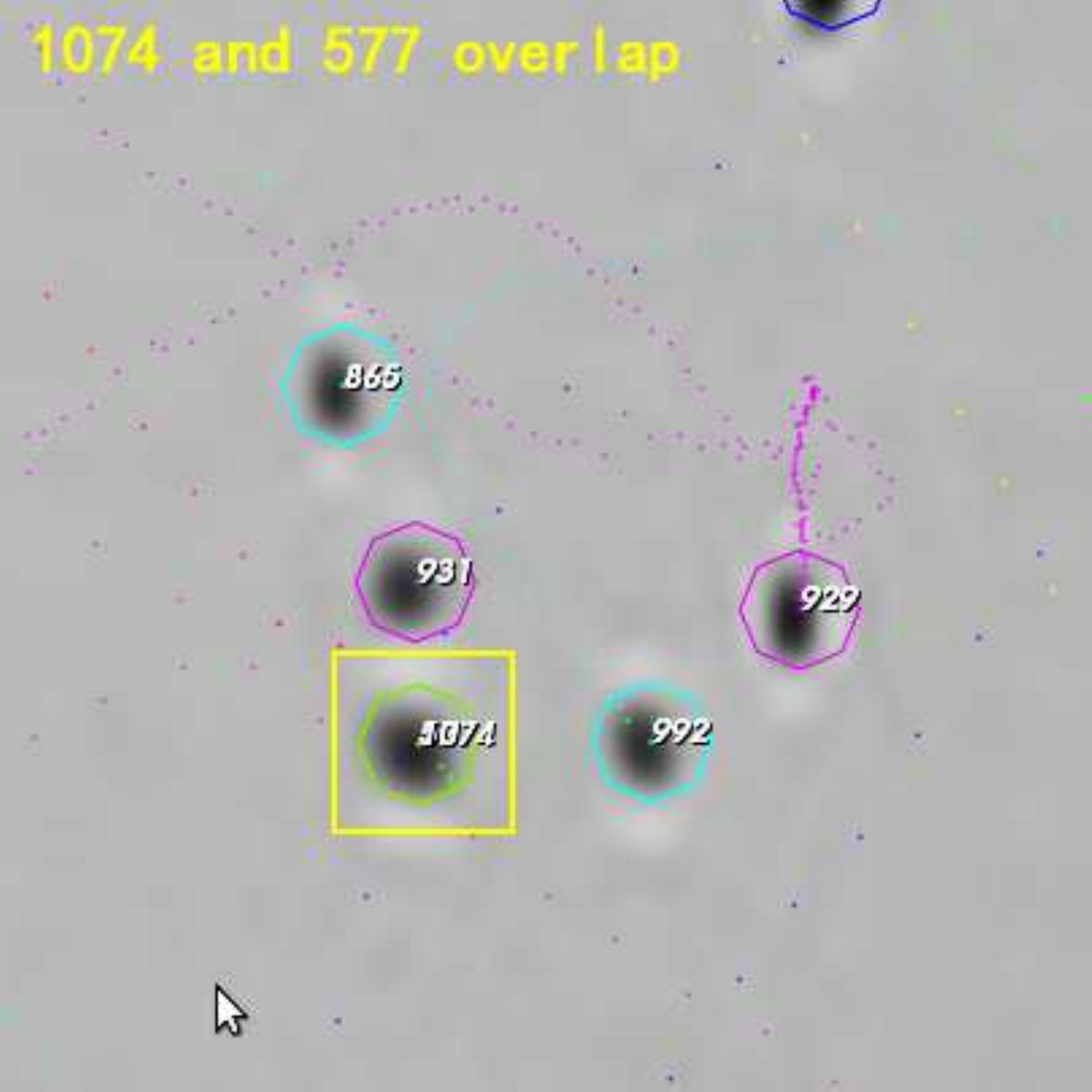}
 \includegraphics[width=0.105\textwidth]{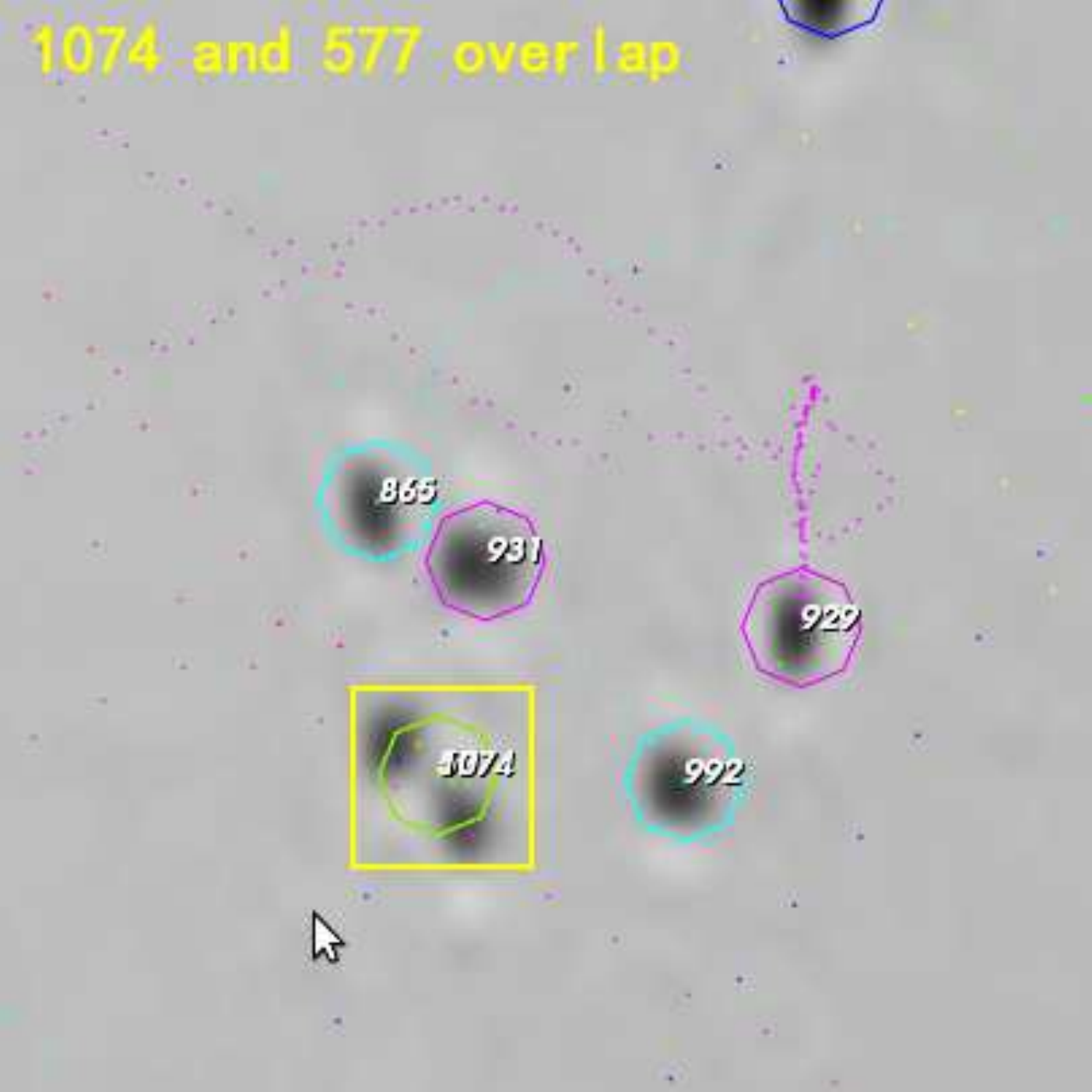}
 \includegraphics[width=0.105\textwidth]{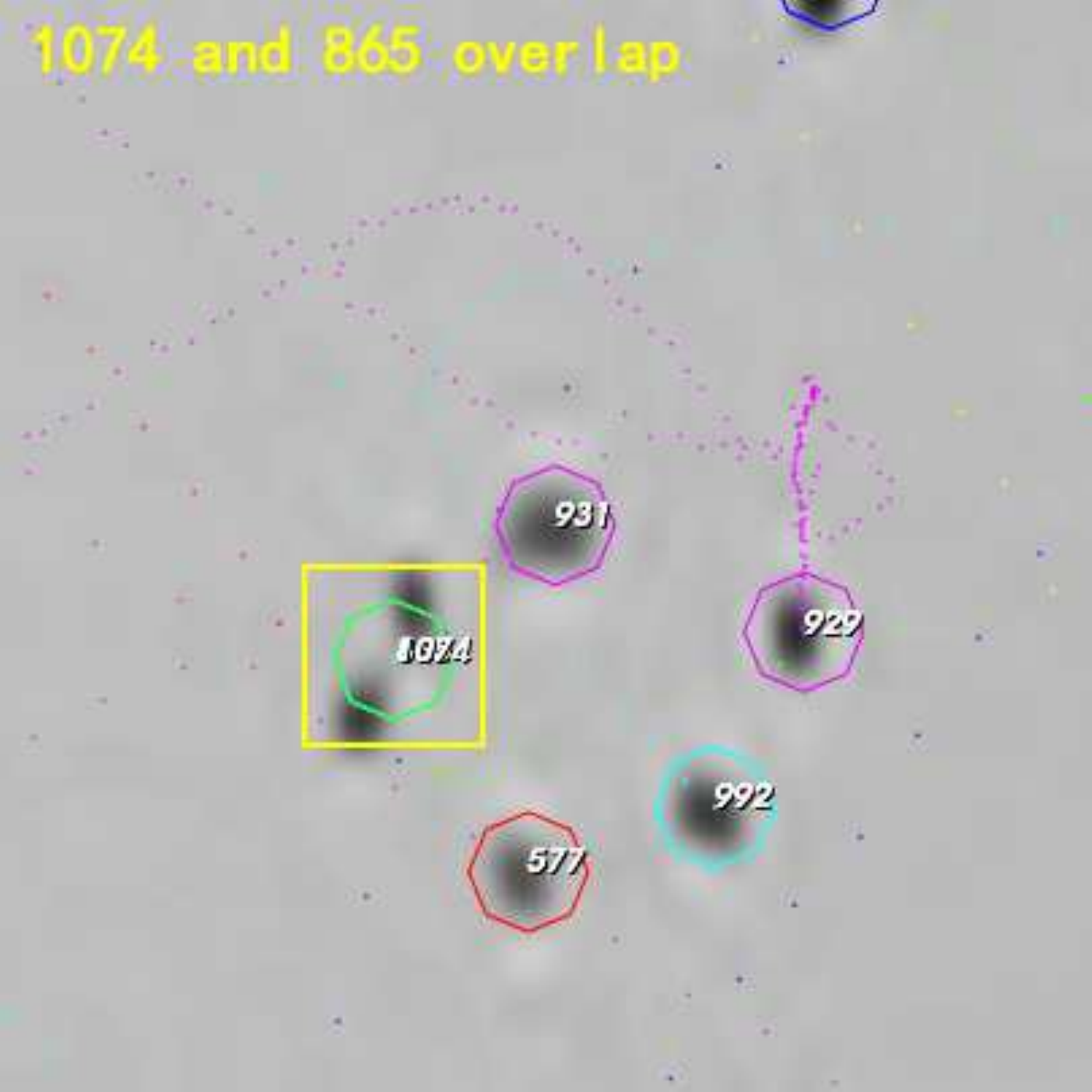}
 \includegraphics[width=0.105\textwidth]{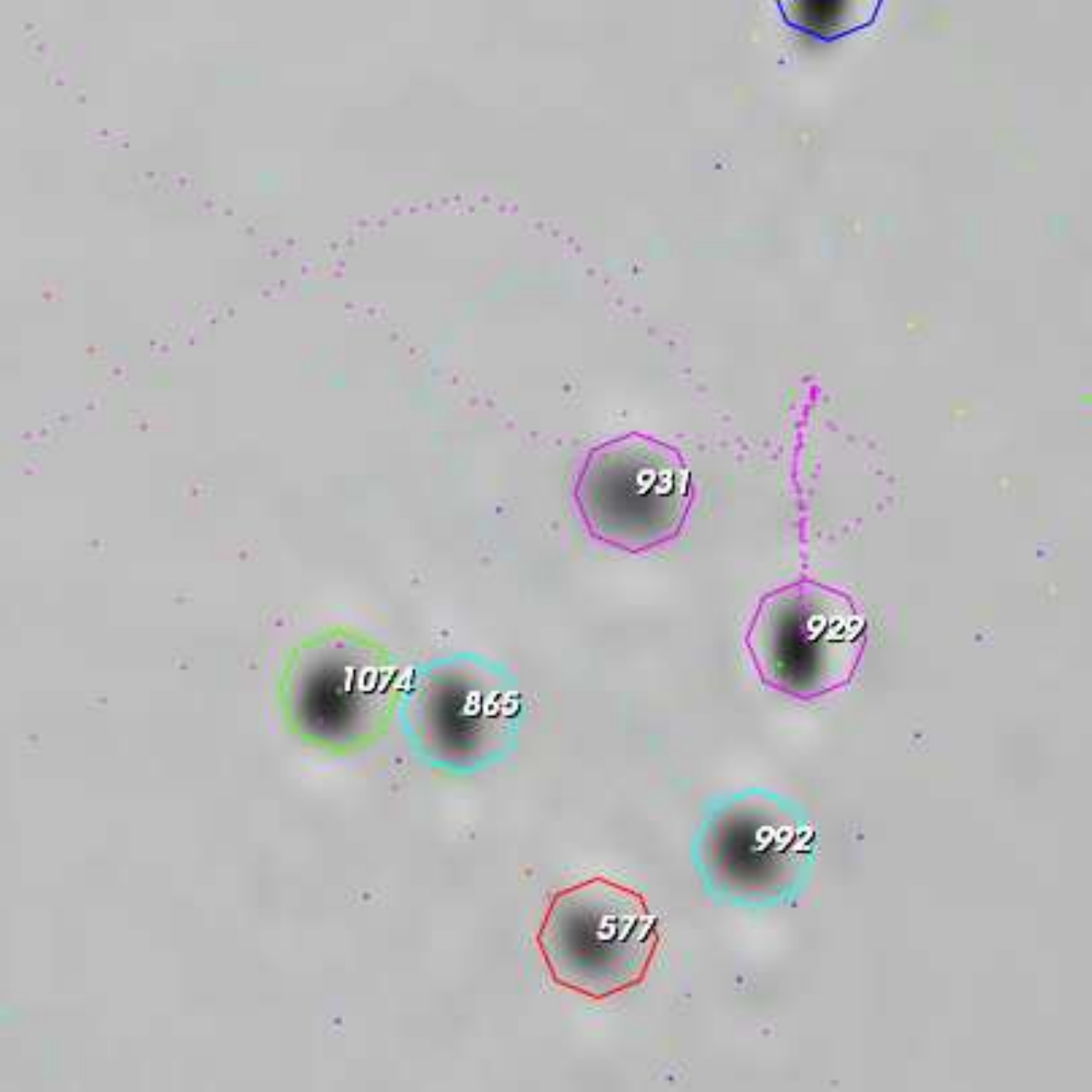}
\includegraphics[width=0.105\textwidth]{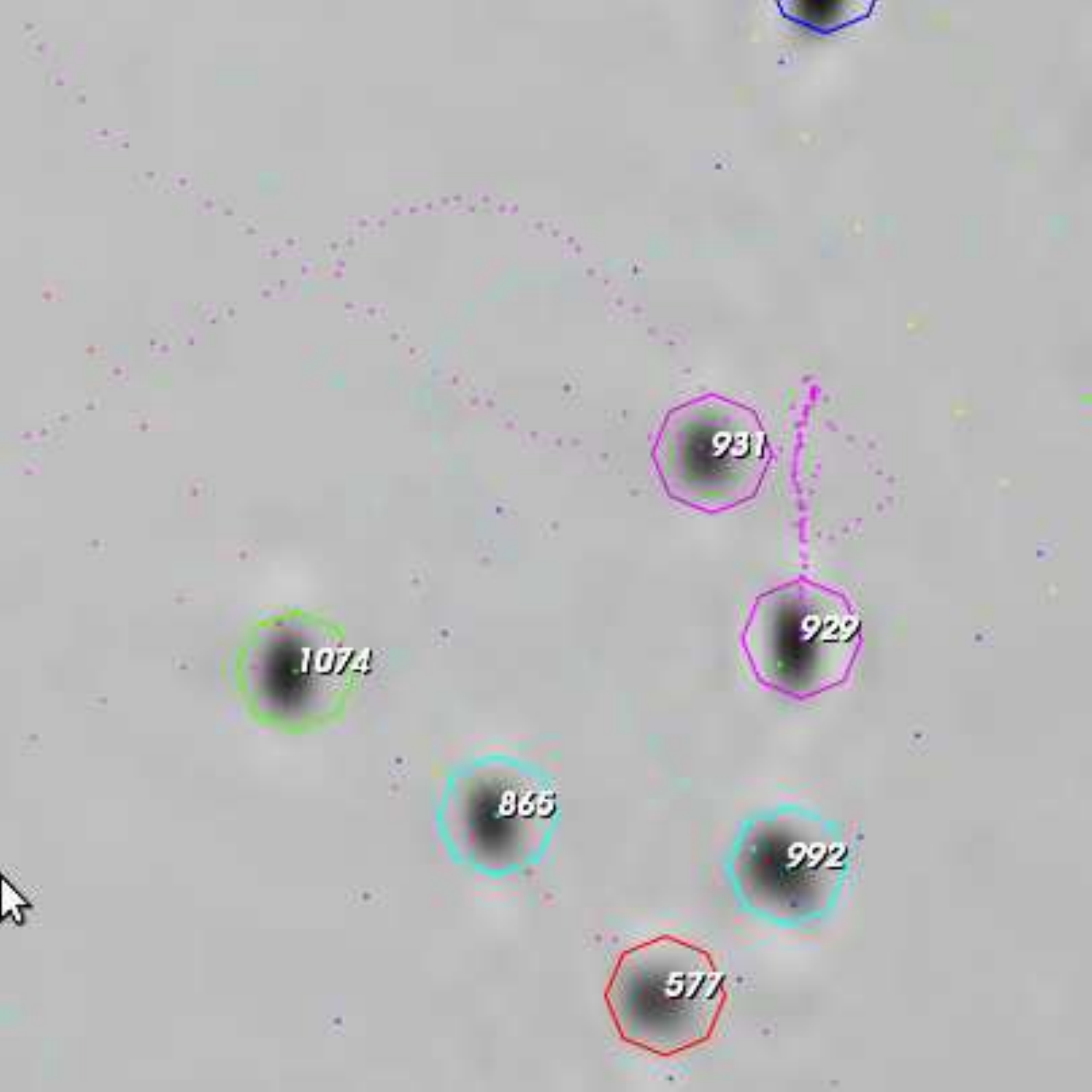}
\end{tabular}
 \caption{The fly trajectories on the 2D image plane resulted from the proposed method, where each fly was labeled with
 a unique number. These flies moved closely to each other
 or overlapped frequently, which causes extreme difficulty to
 establish correct temporal correspondences even by manually
 labeling.} \label{fig:2d_flies}

 \end{figure*}

We tried to establish the 'ground truth' through manually
tracking the flies on image planes and then matching these 2D
trajectories to generate the trajectories of flies, but found
that the task was too difficult even for human vision (\Fig{fig:2d_flies}). We therefore
evaluated the results by viewing the results both in 3D coordinate
system and 2D image planes frame by frame (\Fig{fig:2d_flies}).
Through observation, we found that most of trajectories were
reconstructed correctly and completely. 
We counted the incomplete trajectories that suddenly appeared
or disappeared in the field of view and found that on average about
$2.3\%$ trajectories at each frame were incomplete.

\begin{figure}
\centering \subfigure[Average trajectory
length]{\includegraphics[width =
0.45\textwidth]{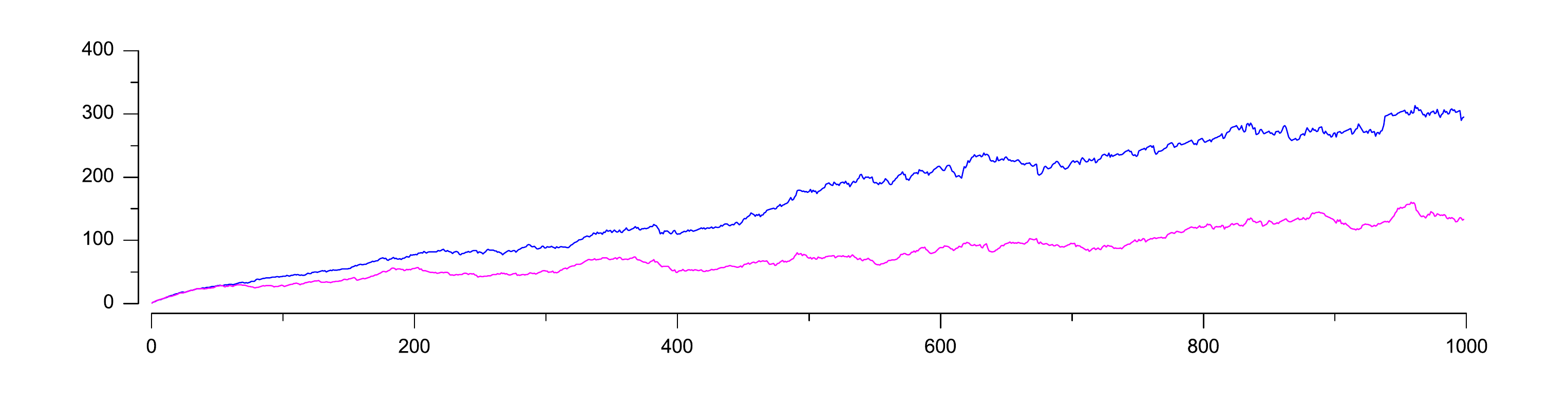}} \subfigure[Number of
flies]{\includegraphics[width = 0.45\textwidth]{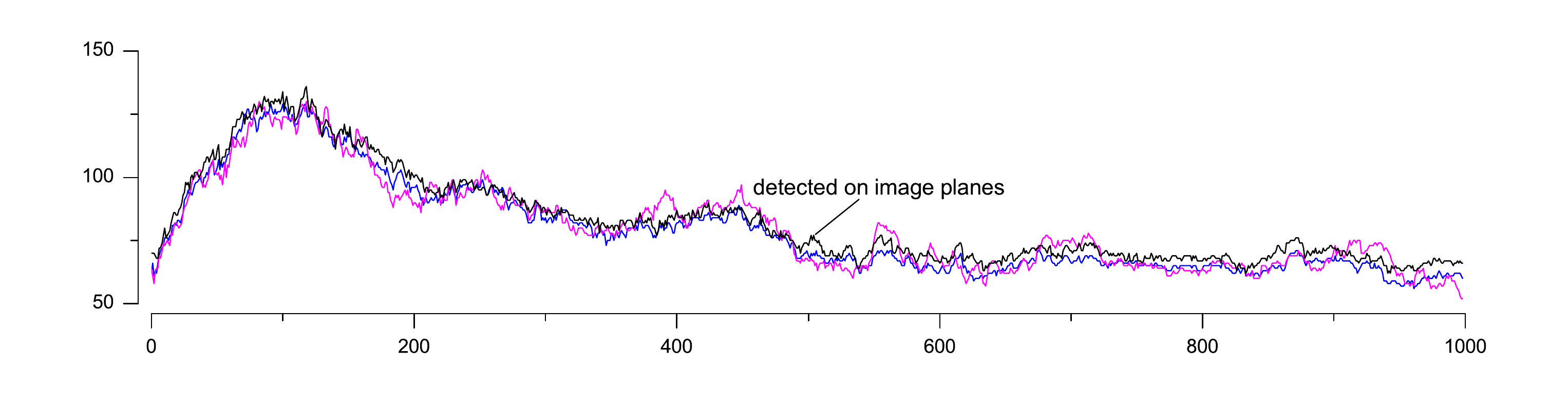}}
\subfigure[Reconstruction consistency ]{
\includegraphics[width = 0.45\textwidth]{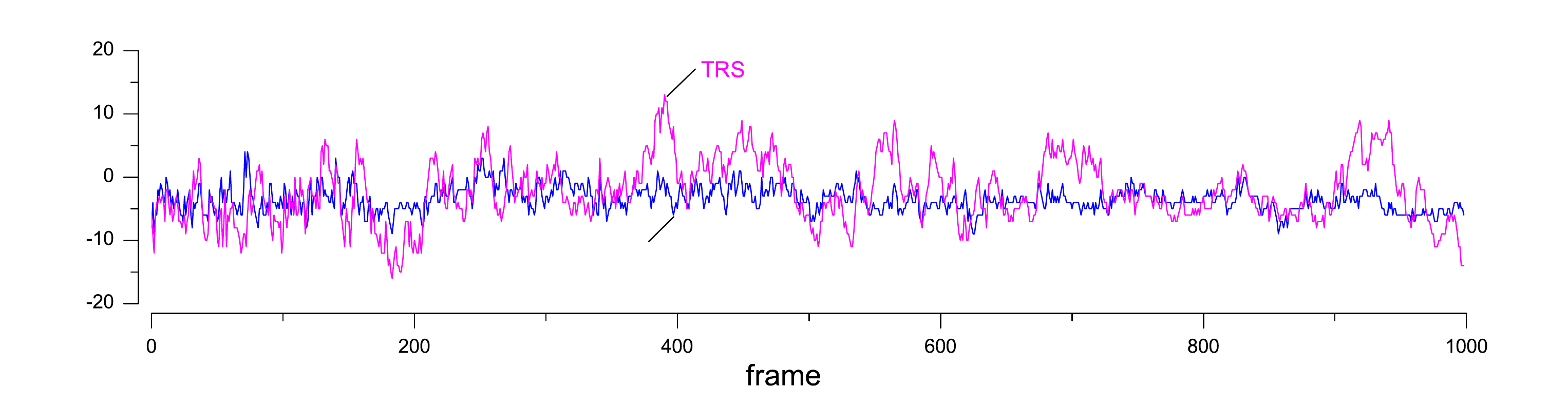}}
 \caption{(a) The average trajectory length at each frame (up to that time step). (b) The number of reconstructed flies
 and the number of detected flies
on the image planes (the smaller one between two cameras). (c) The reconstruction consistency is measured by the
difference between the number of reconstructed 3D flies and the one
detected on image planes. The variance of number difference of the
proposed method is $4.15$, while the one of the TRS method is
$23.98$, indicating the number of flies reconstructed by the
proposed method are more consistent with flies present in video
sequences.
}
 \label{fig:len_num}
 \end{figure}

\begin{figure*}
 \centering
  \begin{tabular}{ccc}
 &
 \begin{minipage}[c]{0.36\textwidth}
 \subfigure[Left view]{\includegraphics[width=
  \textwidth]{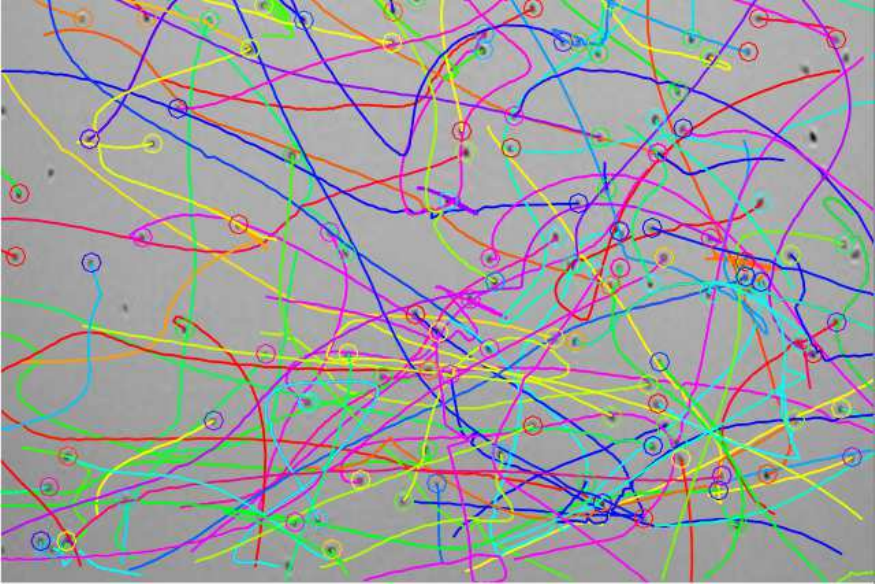}}\\
 \subfigure[Right view]{\includegraphics[width
 =\textwidth]{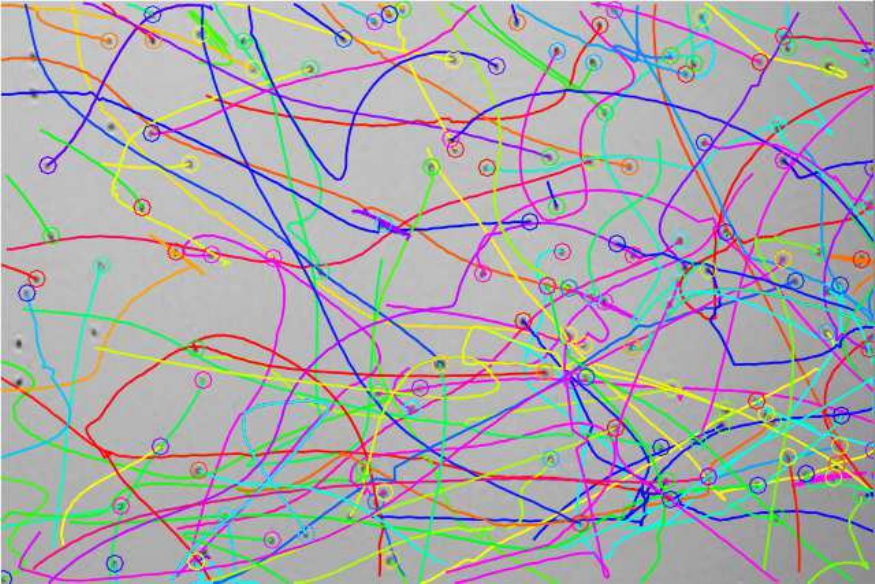}}
 \end{minipage}
 &
 \begin{minipage}[c]{0.535\textwidth}
 \subfigure[The 3D motion trajectories ( The axis unit is $mm$)]{\includegraphics[width
 =\textwidth]{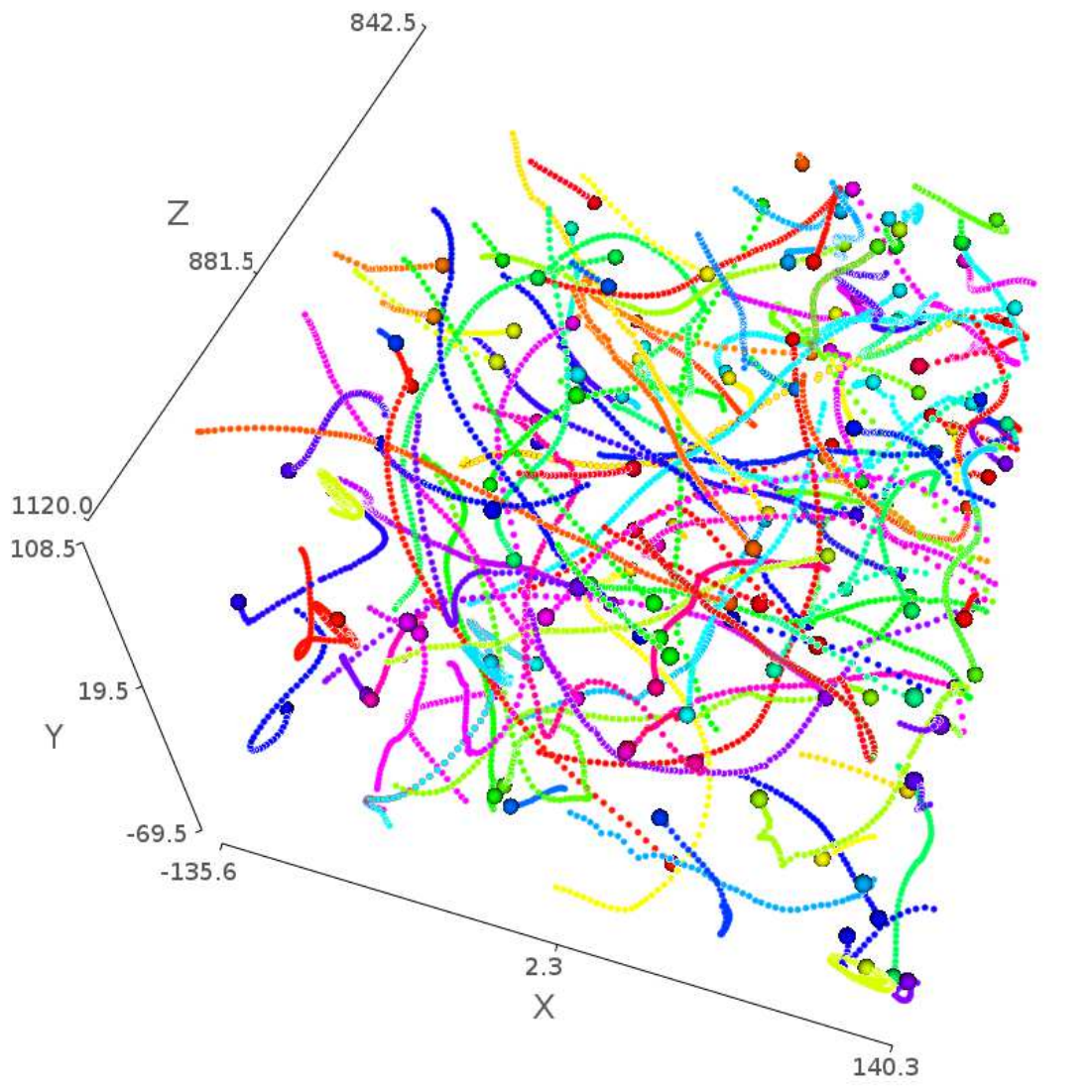}}
 \end{minipage}
 \end{tabular}
  \caption{ The current frame
  here is $96$.  The
  corresponding flies both on image planes and in 3D space are marked
  by the same color.
 }
 \label{fig:frame_by_frame}
 \end{figure*}

 \Fig{fig:len_num}(a) shows the average length of reconstructed
trajectories up to that time at each frame. For comparison, we also gave the result
generated by the TRS method (the best one among the existing methods
as shown in previous sections). We can see that the average
trajectory lengths of the proposed method are much longer than those
of the TRS method, which indicates more trajectories are correctly
reconstructed, as false trajectories seldom have long duration
because of the epipolar motion \cite{du2007rem}. The number of
reconstructed flies and the number of detected flies on the image
plane (the minimum among the two cameras) at each frame are shown
 in \Fig{fig:len_num}(b). The result indicates most flies are successfully
tracked except that a few flies might be captured by only one
camera (such as unmarked flies near image boundaries in
\Fig{fig:frame_by_frame}(a)/(b)). The performances of the TRS method
are also presented
 in \Fig{fig:len_num}. It can be
 seen that the proposed method exhibits a remarkable advantage over
 the TRS method.

\Fig{fig:2d_flies} shows tracking results of the proposed method on
the image plane. The flies moved closely to each other or overlapped
frequently. In such situation, it is extremely difficult to yield
the trajectories of flies by using object tracking techniques on the
2D image plane or even by manually labeling. The proposed method,
however, can easily maintain the identities of flies over time, in
spite of the existence of severe tracking ambiguities.

\begin{figure}
\centering
\includegraphics[width = 0.45\textwidth]{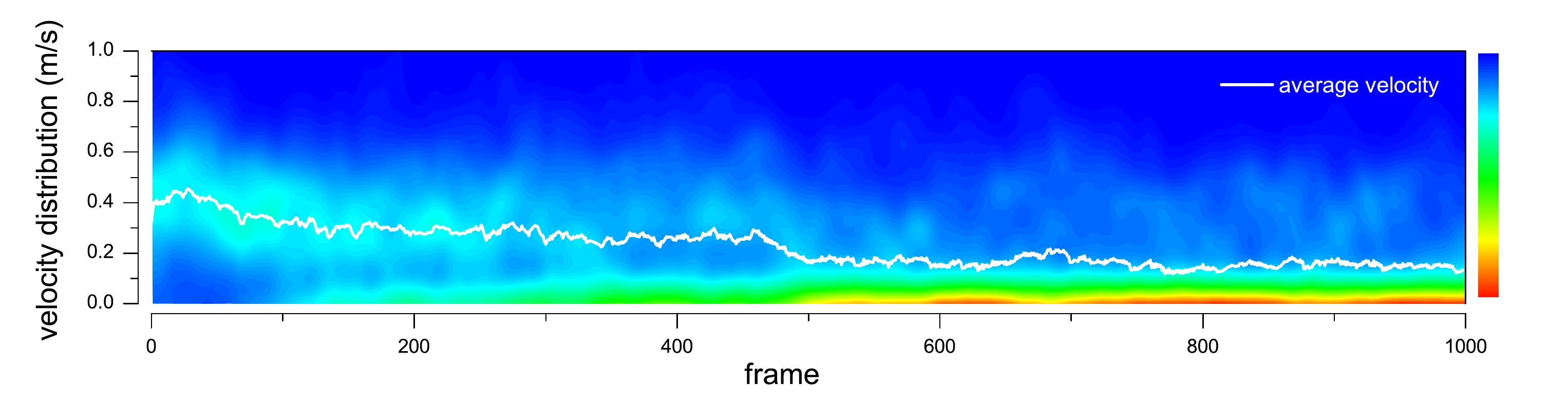}
\caption{The velocity distribution of the flies at each frame.
Notice that the velocities of flies decline when time passes by.
This is because flies tend to land on the wall of the glass box
after a period of flying.} \label{fig:velocity}
\end{figure}

\section{Conclusion}
We have developed a novel approach to measuring the 3D motion trajectories
of large swarm of moving objects of identical appearance by using two
cameras. Unlike the existing methods, the proposed method unifies
tracking and stereo matching into a single cost optimization
problem, mitigating the ambiguities encountered by each of them. Experimental results show clear advantange of the proposed method over the existing methods.

We have used the proposed method to reconstruct 3D
motion trajectories of a large group of flying fruit flies. To our best
knowledge, this is the first time that the 3D motion trajectories of
a large swarm of flying insects are successfully obtained. With these
trajectories, the kinetic information of the flies can be computed.  e.g. We can extract  velocity distribution of the
flying flies at each frame as shown in \Fig{fig:velocity},  and
analyze a single trajectory to study the individual flight behavior
as shown in \Fig{fig:single_trj}. The availability of 3D motion
trajectories of each individual of a flying swarm provides
an opportunity of quantitative analysis for behavior study.
\ifCLASSOPTIONcompsoc
  \section*{Acknowledgments}
\else
  \section*{Acknowledgment}
\fi

The research work presented in this paper is supported by National Natural Science Foundation of China, Grant No. 60875024, Education Commission of Shanghai Municipality Grant No. 10ZZ03, and Science and Technology Commission of Shanghai Municipality, Grant No. 09JC1401500.

\ifCLASSOPTIONcaptionsoff
  \newpage
\fi

\bibliographystyle{IEEEtran}

\end{document}